\documentclass[11pt]{article}
\usepackage{amsmath,amssymb}
\usepackage{amsfonts}   
\usepackage{graphicx} 

\usepackage[toc]{appendix}

\usepackage{enumitem}

\usepackage{booktabs}
\usepackage{bbm}
\usepackage[margin=15pt]{caption}
\usepackage{float}

\usepackage[numbers,compress]{natbib}

 \usepackage{color}
 \usepackage[
       colorlinks=false,
       linkcolor=blue,
       urlcolor=blue,
       filecolor=blue,
       citecolor=red,
       pdfstartview=FitV,
        bookmarksopen=true,
        linktoc=page
       ]{hyperref}

\setlist{nolistsep}

\def\eo{\overset{_{\phantom{.}\circ}}{e}{}}

\def\go{\overset{_{\phantom{.}\circ}}{g}{}}

\def\Do{\overset{_{\phantom{.}\circ}}{D}{}}

\def\Ro{\overset{_{\phantom{.}\circ}}{R}{}}
\def\etao{\overset{_{\phantom{.}\circ}}{\eta}{}}

\newcommand{\fr}{\mathfrak{f}_{\scriptscriptstyle{F \hspace{-0.8mm} R}}}

\newcommand{\cR}{{\cal R}}

\newcommand{\cK}{{\cal K}}

\newcommand{\cX}{{\cal X}}

\newcommand{\cZ}{{\cal Z}}

\def\eql{{~=~}}

\def\bfs#1{{\mathbf #1}}

\newcommand{\sect}[1]{\setcounter{equation}{0}\section{#1}}
\renewcommand{\theequation}{\arabic{section}.\arabic{equation}}

\begin{document}
 
\begin{titlepage}
\hfill DAMTP-2014-55\hspace*{1.7mm}

\hfill  AEI-2014-056\hspace*{0.5mm}
\vspace{1cm}
\begin{center}

{{\Large  \bf  An SO(3)$\times$SO(3) invariant solution of $D=11$ supergravity}} \\

\vskip 1.5cm {Hadi Godazgar$^{\star}$, Mahdi Godazgar$^{\dagger}$, Olaf Kr\"{u}ger$^{\ddagger}$,
\vskip 0.1cm

Hermann Nicolai$^{\S}$ and Krzysztof Pilch$^{\P}$}
\\
{\vskip 0.7cm
$^{\star \dagger}$DAMTP, Centre for Mathematical Sciences,\\
University of Cambridge,\\
Wilberforce Road, Cambridge, \\ CB3 0WA, UK\\
\vskip 0.25cm
$^{\ddagger \S}$Max-Planck-Institut f\"{u}r Gravitationsphysik, \\
Albert-Einstein-Institut,\\
Am M\"{u}hlenberg 1, D-14476 Potsdam, Germany
\vskip 0.25cm
$^{\P}$Department of Physics and Astronomy, \\
University of Southern California, \\
Los Angeles, CA 90089, USA}
{\vskip 0.5cm
$^{\star}$H.M.Godazgar@damtp.cam.ac.uk, $^{\dagger}$M.M.Godazgar@damtp.cam.ac.uk, 
\vskip 0.08cm

$^{\ddagger}$Olaf.Krueger@aei.mpg.de, $^{\S}$Hermann.Nicolai@aei.mpg.de, $^{\P}$pilch@usc.edu}
\end{center}

\vskip 0.35cm

\begin{center}
October 19, 2014
\end{center}

\noindent

\begin{abstract}

We construct a new SO(3)$\times$SO(3) invariant non-supersymmetric solution of the bosonic field equations of $D=11$ supergravity from the corresponding stationary point of maximal gauged $N=8$ supergravity by making use of the non-linear uplift formulae for the metric and the \hbox{3-form} potential. The latter are crucial as this solution appears to be inaccessible to traditional techniques of solving Einstein's field equations, and is arguably the most complicated closed form solution of this type ever found. The solution is also a promising candidate for a stable non-supersymmetric solution of M-theory uplifted from gauged supergravity.  The technique that we present here may be applied more generally to uplift other solutions of gauged supergravity.

\end{abstract}

\end{titlepage}

\setcounter{tocdepth}{2}
\tableofcontents

\newpage

\sect{Introduction}

Kaluza-Klein theory plays an important role as an organising framework in supergravity relating higher and lower-dimensional theories to one another as well as providing a tool by which to derive new theories by dimensional reduction.  Nevertheless, one is confronted with some challenging issues, such as the question of whether a lower-dimensional theory can be obtained from a reduction of a higher-dimensional one, and if so, whether the reduction is consistent.  That is, whether all solutions of the lower-dimensional theory can be mapped onto a subset of the higher-dimensional solutions. How this is done in practice, i.e. how one uplifts solutions to higher dimensions, is yet another level of complication.  Indeed, examples of such results are rare and are mainly confined to truncations with relatively simple scalar sectors.

Eleven-dimensional supergravity compactified on a seven-sphere is one example in which progress has been made; the four-dimensional theory associated with this reduction being maximal SO(8) gauged supergravity.  Recently, an uplift ansatz has been derived for the seven-dimensional components of the 3-form potential in terms of the (pseudo)scalars of the gauged theory \cite{dWN13,testing}.  This complements the uplift ansatz for the seven-dimensional components of the metric given in Ref.~\cite{dWNW}.  Together, these 
ans\"atze give a new method for constructing solutions of $D=11$ supergravity, and
it is the purpose of the present paper to explicitly demonstrate the utility of this new method.
Indeed, without the new uplift formula for the internal flux it is basically 
impossible to construct the solution to be presented in this paper, or to derive any other
solutions of this type that are more complicated than those already in the literature (see for example Refs.~\cite{englert, dWNso7soln, PW, dWNW, Corrado:2001nv}). This is because in all previous examples of solutions corresponding to critical points, the symmetry of the solution reduces the equations of motion to a set of ODEs. In particular, if one obtains the metric via the metric lift ansatz, the equations for the components of the flux field strength are algebraic and 
usually easy to solve.  The analysis becomes even simpler if one has supersymmetry, 
where the ODEs are first order, as is the case for the G$_2$ \cite{dWNW} and SU(3)$\times$U(1) \cite{Corrado:2001nv} solutions. 

The ans\"atze can be applied to obtain a very general class of solutions of $D=11$ supergravity. In particular, they facilitate the uplifting of all stationary points to Freund-Rubin compactifications \cite{freundrubin} with flux, \textit{viz.}
\begin{equation} \label{frwflux}
 {E_M}^A(x,y) = \begin{pmatrix}
                 \Delta^{-1/2}(y) {\eo_{\mu}}{}^{\alpha}(x) & 0 \\[2mm] 0 & {e_m}^a(y)
                \end{pmatrix},
\quad \ F_{MNPQ} = \begin{cases}
                   F_{\mu \nu \rho \sigma} = i \fr \etao_{\mu \nu \rho \sigma} \\[2mm]
                   F_{mnpq} = F_{mnpq}(y), \\[2mm]
                   0,\quad \text{otherwise}
                  \end{cases} \quad \Psi_M=0,
\end{equation}
with the corresponding metric
\begin{equation}
G_{MN} d X^M d X^N = \Delta^{-1} \etao_{\mu \nu} d x^\mu d x^\nu + g_{mn} d y^m d y^n, 
\end{equation}
where $(x^\mu, y^m)$ are coordinates on the four and compact seven-dimensional spacetimes, respectively; ${\eo}_{\mu}{}^{\alpha}(x)$ (with corresponding metric $\etao_{\mu \nu}$) is the vierbein of the maximally symmetric four-dimensional spacetime with corresponding alternating tensor $\etao_{\mu \nu \rho \sigma}$; ${e_m}^a(y)$ (with corresponding metric $g_{mn}$) is the siebenbein of the compact space and $\fr$ is a constant.  In what follows we consider the siebenbein to be that of a deformed round seven-sphere, i.e.
\begin{equation}
 e_m{}^a(y) = \eo_m{}^b(y) S_{b}{}^a(y),
\end{equation}
where $\eo_m{}^a$ (with corresponding metric $\go_{mn}$) corresponds to the siebenbein on a round seven-sphere of inverse radius $m_7$ and the deformation parameter $S$ has determinant $\Delta$,
\begin{equation}\label{DelbyS}
 \Delta(y) = \text{det} S_a{}^b(y).
\end{equation}

The uplift ans\"atze are derived within the context of the SU(8) invariant reformulation of the $D=11$ theory \cite{dWNsu8}, whereby eleven-dimensional fields are decomposed in a $4+7$ split, such that one can loosely talk of them as having external/internal indices.  Note that SU(8) is the local enhanced symmetry obtained in the toroidal reduction of $D=11$ supergravity to four dimensions, with associated global group E$_{7(7)}$ \cite{cremmerjulia}.  Importantly, however, no truncation is assumed and the reformulation remains on-shell equivalent to $D=11$ supergravity \cite{CJS}.  The SU(8) structures in the reformulation are obtained by an analysis of the $D=11$ supersymmetry transformations in such a $4+7$ split, and by the enlargement
of the original SO(7) tangent space symmetry to a full chiral SU(8) symmetry; the R-symmetry of $N=8$ supergravity.

The uplift ans\"atze for the internal metric and flux are derived by comparing the supersymmetry transformations of particular components of the eleven-dimensional fields, namely those with a single ``four-dimensional'' index: the graviphoton $B_{\mu}{}^m$ and $A_{\mu mn}$, which contain the internal metric and 3-form potential components, and the supersymmetry transformation of the associated vectors in four dimensions, which are given in terms of the (pseudo)scalar expectation values.

In this paper, we demonstrate the utility of the uplift ans\"atze by applying them to the only known stable non-supersymmetric solution of the gauged theory \cite{Fischbacher:2010ec,Fischbacherency}: the SO(3)$\times$SO(3) invariant stationary point \cite{Warner84}.  This yields a new solution of $D=11$ supergravity: see equations (\ref{soln:met}, \ref{soln:3form}) and  (\ref{soln:metam}, \ref{soln:3formam}) for the solution in stereographic and ambient coordinates, respectively. This solution, to our knowledge, is the most non-trivial closed form solution of this type ever found (inspection of the explicit formulae in section~\ref{sec:11dsoln} of this paper will probably immediately convince readers of the correctness of this claim). Indeed, the remarkable efficiency of the uplift formulae is clearly demonstrated by the fact that it is significantly simpler to write down the solution than to verify that it does indeed satisfy the $D=11$ equations of motion.

Note that there are many known stable non-supersymmetric compactifications of $D=11$ supergravity of the form $AdS_4\times M_7$ (see e.g.~Ref.~\cite{Page:1984ad}) or indeed $AdS_5\times M_6$ \cite{Dolan:1984kc,Pope:1988xj,Martin:2008pf}, or even purely eleven-dimensional solutions, such as for example, the eleven-dimensional Schwarzschild-Tangherlini solution \cite{Tangherlini:1963bw}.  However, the solution we construct here is the first such solution, as far as we are aware, with non-trivial internal flux and uplifted from maximal gauged supergravity.  While we cannot comment on the eleven-dimensional stability of the solution, the fact that the compactification is stable \cite{Fischbacher:2010ec} in the sense of Breitenlohner-Freedman (BF) \cite{BF} is promising.  Eleven-dimensional stability would be established by demonstrating that the fluctuations associated with higher Kaluza-Klein states also remain above the BF bound.

The SO(3)$\times$SO(3) invariant stationary point is a distinguished solution of the gauged theory. Not only is it the only known stable non-supersymmetric solution, but it also has the most negative value of the cosmological constant of all known stable points and several unstable points \cite{Fischbacherency} and is, therefore, likely \cite{bobev11} to be an attractive IR fixed point for many flows in the world-volume theory on M2-branes \cite{ABJM}.  One example of an RG flow in which this solution is the IR fixed point is that considered in Ref.~\cite{Fischbacher:2010ec}, where the UV fixed point is given by the maximally symmetric SO(8) invariant stationary point \cite{freundrubin,duffpope}.  The study of such RG flows is important in so-called top-down holographic applications to condensed matter systems (see e.g.~Refs.~\cite{popedielec, bobev10}).

The SO(3)$\times$SO(3) invariant solution is an example of a compactification of the form \eqref{frwflux}.  Therefore, the uplift ans\"atze for the metric and internal flux given in Refs.~\cite{dWNW, dWN13} suffice.  In this case, the eleven-dimensional field equations~\footnote{We use the conventions of \cite{dWNsu8}.}
\begin{gather}
 R_{MN} = \textstyle{\frac{1}{72}} g_{MN} F^2_{PQRS} - \textstyle{\frac{1}{6}} F_{MPQR} {F_N}^{PQR}, \label{11dEinstein} \\[7pt]
 E^{-1} \partial_{M} (E F^{MNPQ}) = \textstyle{\frac{\sqrt{2}}{1152}} i \eta^{NPQR_1\ldots R_4 S_1 \ldots S_4} F_{R_1\ldots R_4} F_{S_1\ldots S_4}, \label{11dMaxwell}
\end{gather}
reduce to \cite{dWNW}
\begin{gather}
{R_{\mu}}^{\nu}=\Big(\textstyle{\frac{2}{3}}\, \fr^2 \Delta^4 + \textstyle{\frac{1}{72}} F_{mnpq} F^{mnpq} \Big) \delta_{\mu}^{\nu}, \label{Einseqn1} \\[6pt]
{R_{m}}^{n}= -\textstyle{\frac{1}{6}} F_{mpqr} F^{npqr} + \Big(\textstyle{\frac{1}{72}} F_{pqrs} F^{pqrs} - \textstyle{\frac{1}{3}} \fr^2 \Delta^4 \Big) \delta_{m}^{n}, \label{Einseqn2} \\[6pt]
\Do_q \big(\Delta^{-1} F^{mnpq}\big)=\textstyle{\frac{1}{24} \sqrt{2}} \,\fr \etao^{mnpqrst}F_{qrst} \label{maxwell},
\end{gather}
where $R_{\mu}{}^{\nu}$ and ${R_{m}}^n$ denote components of the eleven-dimensional Ricci tensor $R_{M}{}^{N}$, $\Do_m$ denotes a background covariant derivative and $\etao_{m_1 \dots m_7}$ is the permutation tensor with respect to the metric $\go_{mn}$. All seven-dimensional indices in the equations above are raised with $g^{mn}$, except for $\etao_{m_1 \dots m_7}$, whose indices are raised with $\go^{mn}$.  We parametrise $AdS_4$ and the seven-sphere such that 
\begin{equation}
 \Ro_{\mu \nu} = 3 m_4^2 \go_{\mu \nu},\qquad \Ro_{mn} = -6 m_7^2 \go_{mn}.
 \label{ricciads}
\end{equation}

There are three constants in \eqref{Einseqn1}-\eqref{ricciads}, namely, $m_4$,  $m_7$ and  $\fr$. It is convenient to choose $m_7$ as the overall scale of the solution, since it is simply related to the coupling constant,~$g$, of the $D=4$ theory \cite{dWNconsist87},
\begin{equation}
m_7=\frac{g}{\sqrt 2}\,.
\end{equation}
The remaining two constants are determined by the value of the scalar potential, ${\cal  P}_{cr}=-  P_*\,g^2$,  at the stationary  point, or, equivalently, the cosmological constant of the solution in four dimensions,   
\begin{equation}\label{m4andP}
m_4^2\eql \frac{2P_*}{3}\,m_7^2\,. 
\end{equation}
The value of the $\fr$ parameter can be obtained from the uplift formulae in \cite{dWNconsist87, NP} or the uplift ansatz for the internal components of the 6-form dual \cite{GGN13, KKdual}. In particular, it has been conjectured that the following relation should hold for any stationary point \cite{NP}
\begin{equation}\label{fandP}
\fr\eql \frac{P_*}{\sqrt 2} \,m_7\,.
\end{equation}
However, a general proof of \eqref{fandP} beyond explicit examples remains an open problem.  It is straightforward to verify that for vanishing scalar  fields  one recovers the maximally supersymmetric $AdS_4\times S^7$  Freund-Rubin solution \cite{freundrubin} given by \eqref{ricciads} with 
\begin{equation}\label{}
 m_4= 2 m_7, \qquad \fr = \pm 3 \sqrt{2} m_7 
\end{equation}
 and no internal flux.

The outline of the paper is as follows: in section~\ref{sec:harmonics} we provide the necessary background in order to be able to present the solution without dealing with the technical details.  Then, in section~\ref{sec:invtensors}, we introduce the objects in terms of which we find the solution, which is presented in section~\ref{sec:soln}. For the reader who is simply interested in the solution, and not the technical details of its derivation, section~\ref{sec:overview} is sufficient. 

In section~\ref{sec:defined-quantities}, we state identities satisfied by the SO(3)$\times$SO(3) tensors -- an outline of the derivation of the identities is given in appendix~\ref{app:idproofs}. The metric ansatz gives $\Delta^{-1} g^{mn}$ and some of the identities listed in section~\ref{sec:so7iden} are used to invert this to find the metric, $g_{mn}$, in section~\ref{sec:11dsoln}. Furthermore, the identities are also used to find and simplify the expression for the 3-form potential, $A_{mnp}$, from the flux ansatz in section~\ref{sec:11dsoln}. The majority of the identities are, however, used, in section~\ref{sec:verif-einst-equat}, to verify that the field equations are satisfied. 

We present the SO(3)$\times$SO(3) invariant stationary point of $D=4$ maximal supergravity \cite{Warner84} in section~\ref{sec:gaugedsoln}.  In particular, we recapitulate the scalar profile of the SO(3)$\times$SO(3) invariant stationary point, which is uplifted by means of the ans\"atze, in section~\ref{sec:11dsoln}, to give the internal components of the metric and 3-form potential of the eleven-dimensional solution.  

In section~\ref{sec:verif-einst-equat}, we verify that the solution found in section~\ref{sec:11dsoln} satisfies the $D=11$ supergravity field equations. Given the general arguments that guarantee that the ans\"atze obtained from the uplift formulae solve the equations, this is not strictly necessary. However, we do this in order to demonstrate the full complexity of the solution as well as to give the reader further confidence that the uplift formulae do indeed provide \textit{bona fide} solutions of the $D=11$ equations.  

Finally, in section~\ref{sec:amb} we re-express the eleven-dimensional solution in terms of ambient and local coordinates, which are better adapted to the isometry of the solution than the stereographic coordinates on $S^7$ used in section~\ref{sec:11dsoln}. 

In order to set conventions, we review some basic material, largely contained in Ref.~\cite{dWNsu8}, in appendix~\ref{app:conventions}. For comparison, we list the identities satisfied by the SO(8) and SO(7) tensors for the G$_2$ and SU$(4)^-$ solutions in appendix~\ref{app:comp}. 
In appendix~\ref{app:singleset}, we demonstrate explicitly that the solution can indeed be expressed solely in terms of a single set of (anti-)selfdual SO(8) tensors, as argued in section~\ref{sec:overview}.  In the final appendix, \ref{app:expliciteg}, we give an explicit representation of seven-dimensional $\Gamma$-matrices and an embedding of $\mathbb{R}^4 \oplus \mathbb{R}^4$ in $\mathbb{R}^8$, which is used in section~\ref{sec:amb}. 

\sect{Overview}
\label{sec:overview}

\subsection{The uplift formulae and invariant tensors on $S^7$}
\label{sec:harmonics}

The (pseudo)scalars  of the maximal gauged supergravity in four dimensions parametrise the noncompact coset $\rm E_{7(7)}/SU(8)$. In the unitary gauge,  the  group elements of the coset are given by the scalar 56-bein \cite{so(8)}
\begin{equation}\label{coset}
{\cal V}(x) \eql\exp \left(\begin{matrix}
0 & \phi_{IJKL}(x) \\
\phi^{IJKL}(x)  & 0
\end{matrix}\right)\eql
\left(\begin{matrix}
u_{IJ}{}^{KL}(x)  & v^{IJKL} (x)\\
v_{IJKL} (x)  & u^{IJ}{}_{KL}(x)
\end{matrix}\right)\in E_{7(7)}\,,
\end{equation}
where $\phi^{IJKL} \equiv \phi_{IJKL}^*$ is a complex, selfdual tensor field:
\begin{equation}
\phi^{IJKL} = \frac1{24} \varepsilon^{IJKLMNPQ} \phi_{MNPQ}.
\end{equation} 
The uplift formulae for the internal metric and 3-form potential \cite{dWNW, dWN13} are then written in terms of the 56-bein, ${\cal V}(x) $, and the Killing vectors, $K_m^{IJ}$, and 2-forms, $K_{mn}^{IJ}$, on $S^7$ as follows:\footnote{For conventions and properties of the Killing spinors and tensors, see appendix~\ref{app:conventions}.}
\begin{equation}
   \big(\Delta^{-1} g^{mn}\big)(x,y) = \frac{1}{8} K^{m\,IJ}(y) K^{n\,KL}(y) 
   \Big[\left( u^{MN}{}_{IJ} + v^{MNIJ} \right) 
    \left( u_{MN}{}^{KL} + v_{MNKL} \right) \Big](x), \label{ans:metric} 
\end{equation}
and
\begin{equation}
\big( \Delta^{-1}g^{pq} A_{mnp}\big)(x,y) 
= - \frac{\sqrt{2}}{96} i  K_{mn}^{IJ}(y) K^{q\, KL}(y) 
  \Big[ \left( u^{MN}{}_{IJ} -  v^{MNIJ} \right) \left( u_{MN}{}^{KL} + v_{MNKL} \right) \Big](x).
 \label{ans:flux}
\end{equation}
In writing these and similar formulae we will adopt and apply the following convention consistently
throughout this paper:\\[1mm]
\noindent
{\em The raising or lowering of indices on any geometric object on  $S^7$, is always done by means of the round $S^7$ metric $\mathring g_{mn}$ and its inverse.  By contrast, to raise or lower indices on the physical fields of D = 11 supergravity (as they appear for instance in (\ref{11dEinstein}) and (\ref{11dMaxwell})), we always employ the full metric $g_{mn}$ and its inverse.}

This means, in particular, that on the right hand side of the above equations we have
$K^{mIJ}\equiv\go^{mn}K_n^{IJ}$ and so on.

The full metric $g_{mn}(x,y)$ is then obtained by inverting and peeling of the
determinant factor using
\begin{equation}\label{warpbla}
\Delta^{-9}\eql \det(\Delta^{-1}g^{mn}\go_{np})\,.
\end{equation}
For the  3-form field, $A_{mnp}(x,y)$, one must then insert the result for the 
densitised metric, $\Delta g_{qr}$, on the right hand side of \eqref{ans:flux}. 

Formulae \eqref{ans:metric} and \eqref{ans:flux} are off-shell in the sense that they  
give the internal metric, $g_{mn}$, and the 3-form potential, $A_{mnp}$,  for  {\it any}  
configuration of the scalar fields  of the maximal gauged supergravity embedded in eleven-dimensional supergravity.  In particular, note that 
the full antisymmetry of $A_{mnp}$ in \eqref{ans:flux} is not manifest, but can be established by 
means of the E$_{7(7)}$ properties of the 56-bein $\mathcal{V}$, and is thus independent of whether the equations of motion are satisfied or not \cite{dWN13}. 

The main task is thus to construct, from a given scalar field configuration $\phi^{IJKL}(x)$,
the geometric quantities $g_{mn}(x,y)$ and $A_{mnp}(x,y)$. To gain a better
perspective on this problem, let us first discuss the construction in a more general context 
before we specialise to SO(3)$\times$SO(3) symmetric configurations below.
For the most general configuration that has no symmetries at all the scalar field configuration would of course 
involve the full set of 35 scalars and  35 pseudoscalars. However, we are here interested 
in specific configurations preserving some symmetry, for which we can restrict attention to~\footnote{Of course, at the stationary point, we can group all scalars and pseudoscalars into single SO(8) invariant objects with associated SO(7) tensors, defined in an analogous manner to those defined in \eqref{xir} and \eqref{Ss}.  In this case, one is guaranteed that the solution may be written solely in terms of these reduced set of SO(7) tensors.  However, the result will not, in general, take a `nice' form (see appendix~\ref{app:singleset} for a demonstration of this for the SO(3)$\times$SO(3) invariant solution). }
\begin{equation}
\phi^{IJKL}(x) = \sum_r \lambda^{(r)}(x) \Phi^{(r)}_{IJKL} +
                            i \sum_s \mu^{(s)}(x) \Psi^{(s)}_{IJKL}
\end{equation}
where $\big\{ \Phi^{(r)}_{IJKL} \big\}$ and $\big\{ \Psi^{(s)}_{IJKL}\big\}$ form a basis of
{\em invariant} real selfdual and real anti-selfdual 4-forms (when we are dealing with
real tensors the position of the indices $I,J,...$ does not matter). If one is looking for 
stationary points preserving a given symmetry, the scalar manifold is accordingly 
parametrised by coordinates $\big\{\lambda^{(r)}, \mu^{(s)} \big\}$. 
Simple examples of invariant 4-forms (for which the labels $r$ and $s$ are not needed) are
\begin{align*}
\Phi_{IJKL} &= C^+_{IJKL}  \; ,   &\Psi_{IJKL} & = 0
\qquad &\mbox{for SO(7)$^+$ symmetry};  \nonumber\\[2mm]
\Phi_{IJKL} &= 0  \; , &\Psi_{IJKL} &=  C^-_{IJKL}
\qquad &\mbox{for SO(7)$^-$ symmetry};  \nonumber\\[2mm]
\Phi_{IJKL} &= C^+_{IJKL}  \; , &\Psi_{IJKL} &=  C^-_{IJKL}
\qquad &\mbox{for G$_2$ symmetry}.  
\end{align*}
For the SO(3)$\times$SO(3) solution we are about to construct, there are two invariant 
selfdual and two invariant anti-selfdual 4-forms, which are given in (\ref{so8tensors}) below.
In order to rewrite the solution in terms of geometric objects adapted to the
(deformed) $S^7$ geometry, we define a set of invariant tensors via
\begin{equation}\label{xir}
\xi^{(r)}_m = \frac1{16} \Phi^{(r)}_{IJKL} K_{mn}^{IJ} K^{n\,KL} \;, \quad
\xi^{(r)}_{mn} = - \frac1{16} \Phi^{(r)}_{IJKL} K_m^{IJ} K_n^{KL} \; , \quad
\xi^{(r)} = \mathring g^{mn} \xi^{(r)}_{mn}
\end{equation}
for the scalars, and
\begin{equation}\label{Ss}
S^{(s)}_{mnp} = \frac1{16} \Psi^{(s)}_{IJKL} K_{mn}^{IJ} K_p^{KL}
\end{equation}
for the pseudoscalars. By virtue of their definition and the (anti-)selfduality properties of 
the invariant 4-forms, these tensors satisfy the relations
\begin{eqnarray}
\Do_m \xi &=& 2 m_7 \xi_m, \hspace{40mm} \Do_m \xi_n^{(r)} = 6 m_7 \, \xi^{(r)}_{mn} - 2 m_7 \, \xi^{(r)} \mathring g_{mn}, \nonumber\\[2mm]
\Do_m \xi_{np}^{(r)} &=& \frac13 m_7 \Big( \go_{np} \xi_m^{(r)} - \go^{\phantom{(r)}}_{m(n} \xi^{(r)} _{p)} \Big), \qquad
\Do_m S^{(s)}_{npq} = \frac16 m_7 \mathring \eta_{mnpq}{}^{rst} S^{(s)}_{rst}
\end{eqnarray}
for all $r$ and $s$. Furthermore, we have the inversion formulae
\begin{equation}\label{invform}
\begin{split}
\Phi^{(r)}_{IJKL} & \eql \frac{1}{6}\xi^{(r)} K_m^{[IJ} K^{m \, KL]} - 
\frac{3}{2} \xi^{(r)\, mn} K_m^{[IJ} K_n^{KL]} + \frac{1}{12} \xi^{(r)\,m} K_{mn}^{[IJ} K^{n \, KL]}\,,\\[2mm]
\Psi^{(s)}_{IJKL} & \eql \frac{1}{2}  S^{(s)}_{mnp} \, K^{mn[IJ}  K^{p\, KL]}\,,
\end{split}
\end{equation}
which are, again, valid separately for all $r$ and $s$.

Now, for any specific set of invariant 4-forms we will need further identities. First, such identities are needed to perform the exponentiation required for 
the calculation of $u^{IJ}{}_{KL}, v^{IJKL}$ and their complex conjugates in (\ref{coset}).
Second, we need these identities to solve the uplift formulae for $g_{mn}$ and $A_{mnp}$
and to bring the resulting expressions into a manageable form. 

The simplest examples, again, are provided by the SO(7)$^\pm$ and 
G$_2$ solutions for which the invariant 4-forms $C^\pm_{IJKL}$ obey
\begin{equation}
C^\pm_{IJMN} C^\pm_{MNKL} = 12 \delta^{IJ}_{KL} \pm 4 C^\pm_{IJKL},
\end{equation}
i.e. their contractions either reproduce the same 4-forms or give the identity.
The general case is more complicated because any product of 4-forms may produce new
invariant tensors that are not 4-forms. The simplest example here is the G$_2$ solution
that depends on both $C^+_{IJKL}$ and $C^-_{IJKL}$, as well as the product
$C^+_{IJMN}  C^-_{MNKL}$, which defines a new invariant tensor (which is not a 4-form);
this object then completes the list of G$_2$ invariant tensors. 
A more complicated example is the tensor $F_{IJ}$ defined in (\ref{Ftens}) 
and further invariant objects for the SO(3)$\times$SO(3) solution. 
Consequently we will need to evaluate products such as 
\begin{equation}
\Phi^{(r)}_{IJMN} \Phi^{(r')}_{MNKL} \; , \;\;
\Phi^{(r)}_{IJMN} \Psi^{(s)}_{MNKL} \; , \;\;
\Phi^{(r)}_{IJMN} \Psi^{(s)}_{MNPQ} \Phi^{(r')}_{PQKL} \; , 
\quad  etc.
\end{equation}
and either reduce them to previously defined expressions or add them as new objects
to the list of invariant tensors. The procedure stops when all products or contractions
reproduce objects already contained in the list; exploiting all such identities should enable
us to compute $u^{IJ}{}_{KL}$ and $v^{IJKL}$ in a closed form.

Furthermore, as we will explain below in much detail for the SO(3)$\times$SO(3) case,
the identities satisfied by the above invariants entail a corresponding hierarchy of 
identities  for the geometric tensors introduced in (\ref{xir}) and (\ref{Ss}). The main
use of these identities will be in carrying out the inversion required to derive the
metric and 3-form from the uplift formulae (\ref{ans:metric}) and (\ref{ans:flux}) and in
bringing the resulting expressions into a manageable form. This last step is necessary
for the verification of the $D=11$ field equations which would otherwise be unmanageably
complicated.

Before proceeding let us comment on another point. In Kaluza-Klein theory one
is usually interested in calculating the mass spectrum of a given compactification,
and the massless states in particular. 
This requires a linearised expansion  of  the metric \eqref{ans:metric} and the 
3-form potential \eqref{ans:flux} in the scalar fluctuations  around a given vacuum. 
For the maximally symmetric $S^7$ compactification we thus have \cite{duffpope, Biran:1982eg}
 \begin{eqnarray}\label{harexp}
g_{mn}(x,y)\eql \go_{mn} (y) +
\sum A_{IJKL} (x)\, {\cal Y}^{IJKL}_{mn}(y)+\ldots \,, \nonumber \\[2mm] 
A_{mnp}(x,y) \eql \sum  B_{IJKL}(x)  \, {\cal Y}^{IJKL}_{mnp}(y)+\ldots \,,
\end{eqnarray}
where $A_{IJKL}$ and $B_{IJKL}$ are the 35 scalar and 35 pseudoscalar fields
of $N=8$ supergravity ($\phi^{IJKL} = A^{IJKL} + iB^{IJKL}$),
and, where the ellipses denote massive modes. The corresponding 
eigenmodes have been known for a long time \cite{duffpope, Biran:1982eg}
\begin{eqnarray}
 {\cal Y}^{IJKL}_{mn}(y) &=& K_m^{[IJ} K_n^{KL]} -
                \frac19 \go_{mn} K^{p[IJ} K_p^{KL]}, \nonumber\\[2mm]
{\cal Y}^{IJKL}_{mnp}(y) &=& K_{[mn}^{[IJ} K_{p]}^{KL]}.
\end{eqnarray}
The formulae (\ref{ans:metric}) and (\ref{ans:flux}) are thus the {\em consistent non-linear
extensions of the above formulae} (it is straightforward to check that the linearised
formulae follow directly from (\ref{ans:metric}) and (\ref{ans:flux}) by expanding the 
latter to first order in the scalar and pseudoscalar fields). One can therefore ask whether
it is possible to directly `exponentiate' the formulae (\ref{harexp}). The above discussion
shows that this is indeed possible for restricted configurations if one has enough
tensor identities at hand.

\subsection{Invariant tensors for the  SO(3)$\times$SO(3)  solution}
\label{sec:invtensors}

The SO(3)$\times$SO(3) subgroup of $\rm SO(8)$, which is the symmetry of the stable stationary point in maximal gauged supergravity, 
is defined by the following branchings of the three fundamental representations:
\begin{equation}\label{}
\bfs 8_v~\longrightarrow~ (\bfs 3,\bfs 1)+(\bfs 1,\bfs 3)+2\times(\bfs 1,\bfs 1)\,,\qquad
\bfs 8_{s,c}~\longrightarrow~2\,\times (\bfs 2,\bfs 2)\,.
\end{equation}
In the conventions that we are using, the eight gravitini, $\psi^I$, and the Killing spinors, $\eta^I$, on $S^7$, transform under $\bfs 8_v$. We choose the two $\rm SO(3)$ groups to act on the subspaces defined by $I=1,2,3$ and $I = 6,7,8$, respectively. Then the four invariant noncompact generators of $E_{7(7)}$ are given by the tensors 
\begin{equation}\label{so8tensors}
\begin{split}
  Y^+_{IJKL} &= 4!  \left(\delta^{1234}_{IJKL} + \delta^{5678}_{IJKL} \right), \qquad
  Y^-_{IJKL} = 4!  \left(\delta^{1235}_{IJKL} + \delta^{4678}_{IJKL} \right),  \\[2mm]
  Z^-_{IJKL} &= 4!  \left(\delta^{1234}_{IJKL} - \delta^{5678}_{IJKL} \right), \qquad
  Z^+_{IJKL} = 4!  \left(\delta^{1235}_{IJKL} - \delta^{4678}_{IJKL} \right), 
\end{split}\end{equation}
where $Y^+_{IJKL}$ and $Z^+_{IJKL}$ are selfdual, while $Y^-_{IJKL}$ and $Z^-_{IJKL}$ are anti-selfdual. In section~\ref{sec:11dsoln}, we show that the simplest and most symmetric form of the solution is obtained in terms of the following invariants defined by these tensors:~\footnote{Cf. definitions \eqref{xir} and \eqref{Ss}.} 
 \begin{gather}
  \xi_m =\; \frac{1}{16} Y^+_{IJKL} K_{mn}^{IJ} K^{n \, KL}, \quad
  \xi_{mn} =\; -\frac{1}{16} Y^+_{IJKL} K_{m}^{IJ} K_n^{KL}, \quad
  \xi =\; \mathring g^{mn} \xi_{mn},\notag \\[2mm]
  \zeta_m =\; \frac{1}{16} Z^+_{IJKL} K_{mn}^{IJ} K^{n \, KL}, \quad
  \zeta_{mn} =\; -\frac{1}{16} Z^+_{IJKL} K_{m}^{IJ} K_n^{KL}, \quad
  \zeta =\; \mathring g^{mn} \zeta_{mn},\notag \\[2mm]
  S_{mnp} =\; \frac{1}{16} Y^-_{IJKL} K_{[mn}^{IJ} K_{p]}^{KL},\qquad  
  T_{mnp} =\; \frac{1}{16} Z^-_{IJKL} K_{[mn}^{IJ} K_{p]}^{KL}, \label{so7tensors}
\end{gather}
as well as two additional tensors
\begin{equation}\label{Ftens}
F_{m}\eql F_{IJ}K_m^{IJ}\,,\qquad F_{mn}\eql F_{IJ}K_{mn}^{IJ}\,,\qquad  F_{IJ}\eql\delta^{45}_{IJ}\,,
\end{equation}
which  satisfy (see section~\ref{sec:defined-quantities})
\begin{equation}\label{}
F_m\eql \frac{1}{18}\, \mathring \eta_{mnpqrst} S^{npq} T^{rst}\,,\qquad \Do_mF_n\eql - m_7\,F_{mn}\,.
\end{equation}
Note that, as emphasised before, the objects defined in \eqref{so7tensors} and \eqref{Ftens} belong to $S^7$. Hence, their indices are raised and lowered with $\mathring g_{m n}$ and its inverse, for instance $\xi^{m n} \equiv \mathring g^{m p} \mathring g^{n q} \xi_{p q}$.

 \subsection{The solution}
\label{sec:soln}

We are now in a position to state the main result of this paper, which is an explicit uplift of the solution at the $\rm SO(3)\times SO(3)$ stationary point of the scalar potential written in terms of the geometric quantities introduced above.  The solution  below is presented in its simplest and the most symmetric form. We refer the reader to section~\ref{sec:11dsoln} for a more general form of the solution which, in particular,  includes an additional parameter, $\alpha$, corresponding to an accidental $\rm U(1)$ symmetry of the potential. The solution below is for $\alpha=-\pi/4$.
 
The internal metric  of  the uplifted solution is 
\begin{equation}\label{soln:met}
g_{mn}\eql \frac{\Delta ^2}{18}\big[ (\mathcal{X}^2 + \mathcal{Z}^2) \mathring g_{mn} - 12
  (\mathcal{X} \xi_{mn} + \mathcal{Z} \zeta_{mn}) + 2 f_m f_n \big], 
\end{equation}
where
\begin{equation}\label{}
f_m\eql 6\,F_m-\sqrt 5\,\xi_m+\sqrt 5\,\zeta_m\,.\\
\end{equation}
The  3-form flux is
\begin{equation}\label{soln:3form}
\begin{split}
A_{mnp} \eql \frac{\Delta^3}{18\sqrt 2} \,\Big[ &
\Big(\frac{\sqrt 5}{3}\,\cZ(\cX-\cZ)-\cX-5\cZ\Big)\,S_{mnp} +
\Big(\frac{\sqrt 5}{3}\,\cX(\cX-\cZ)+5\cX+\cZ\Big)\,T_{mnp}
\\
&+ \frac{1}{27}  \mathring \eta_{mnpqrst} \left( \mathcal{Z} \xi^q - \mathcal{X} \zeta^q \right)
  \left( \mathcal{Z} S^{rst} + \mathcal{X} T^{rst}
  \right)-\frac{4}{3} \left( \mathcal{X} \xi_{[m} + \mathcal{Z} \zeta_{[m}\right) S_{np]q} \xi^q \,
\Big]\,,
\end{split}
\end{equation}
where the warp factor, $\Delta$,  is given by
\begin{equation}\label{}
\Delta^3 \eql \frac{36}{\mathcal{X}^2 + 10 \mathcal{X} \mathcal{Z} + \mathcal{Z}^2 }\,,
\label{det}
\end{equation}
with
\begin{equation}\label{}
\cX\eql 2\,\xi-3\sqrt 5\,,\qquad \cZ\eql  2\,\zeta-3\sqrt 5\,.
\end{equation}
The  solution is now complete modulo two constants, which as discussed in the introduction, are determined by the value of the potential, $P_*$, at the stationary point using \eqref{m4andP} and \eqref{fandP}.  For the SO(3)$\times$SO(3) point,  $P_*=14$.  Hence,
\begin{equation}
  m_4^2 = \frac{28}{3} m_7^2\,,\qquad \fr = 7 \sqrt{2} m_7\,.
\label{m4fr}
\end{equation}
In particular, the fact that the value of $\fr$ given above, as determined by equation \eqref{fandP}, is consistent with a solution of the equations of motion is further evidence for the validity of this conjectured relation \eqref{fandP} between $\fr$ and the potential.   The remaining constant, $m_7$, sets the overall scale of the solution.

One should note that the metric  and the 3-form potential  in \eqref{soln:met} and \eqref{soln:3form} are obtained by an application of the identities derived in section~\ref{sec:defined-quantities} to simplify the  ``raw'' expressions that follow from the uplift formulae. We refer the reader to section~\ref{sec:11dsoln} for details of the derivation and to section~\ref{sec:amb} for another form of the solution in which the geometry of the internal space is perhaps more transparent.

\sect{Identities for SO(3)$\times$SO(3) invariants}
\label{sec:defined-quantities}

In this section, we present in a systematic way a set of identities for the geometric objects 
\begin{equation}\label{ourtens}
\xi\,,\quad \zeta\,,\quad \xi_m\,,\quad \zeta_m\,,\quad \xi_{mn}\,,\quad \zeta_{mn}\,,\quad S_{mnp}\,,\quad T_{mnp}\,,\quad F_m\,,\quad F_{mn}\,,
\end{equation}
defined in \eqref{so7tensors} and \eqref{Ftens}. These identities are crucial for the discussion in subsequent sections, in particular, we need them to derive, in section~\ref{sec:11dsoln}, the simplified form of the solution in \eqref{soln:met} and \eqref{soln:3form} and  
to verify that the equations of motion are satisfied in section~\ref{sec:verif-einst-equat}. 
They are also of interest on their own  as the starting point for identifying the SO(3)$\times$SO(3) geometry underlying  our solution. Such an identification would allow one  to construct a large class of new solutions  in which the underlying internal manifold is not necessarily the  round seven-sphere in much the same way as is done when extending the  $\rm SU(4)^-$ solution \cite{PW} to arbitrary Sasaki-Einstein manifolds \cite{Pope:1984jj, PW, Gauntlett:2009zw}.

The identities we are looking for fall into two broad categories: (i)  generic identities, which are proved using only the (anti-)selfduality property of the underlying $\rm SO(8)$ tensors and  properties of the Killing vectors/spinors;~\footnote{All identities in section~\ref{sec:harmonics} fall into this category.} (ii) identities specific to the objects \eqref{ourtens}. These are proved by exploiting the concrete SO(3)$\times$SO(3) invariant form of the  $\rm SO(8)$ tensors $Y^\pm_{IJKL}$, $Z^\pm_{IJKL}$ and $F_{IJ}$ defined in \eqref{so8tensors} and \eqref{Ftens}.

\subsection{Generic identities}
\label{sec:genso7id}

The identities in this section follow from the particular dependence of the SO(7) tensors \eqref{ourtens} defined in \eqref{so7tensors} and \eqref{Ftens} on the  Killing vectors/spinors. They do not require  specific knowledge of how the underlying SO(8) tensors are defined. We refer the reader to Refs.~\cite{dWNso7soln, parallel, rel411, testing} for proofs and further details. 

Equations  \eqref{so7tensors} and \eqref{Ftens} can be inverted using the completeness property of the $\Gamma$-matrices. This yields, cf.\ \eqref{invform},
\begin{gather}
Y^+_{IJKL} = \frac{1}{6}\xi K_m^{[IJ} K^{m \, KL]} - \frac{3}{2} \xi^{mn} K_m^{[IJ} K_n^{KL]} + \frac{1}{12} \xi^m K_{mn}^{[IJ} K^{n \, KL]}, \quad
Y^-_{IJKL} = \frac{1}{2} S^{mnp} \, K_{mn}^{[IJ}  K_{p}^{KL]}, \notag \\[2mm] 
Z^+_{IJKL} = \frac{1}{6}\zeta K_m^{[IJ} K^{m \, KL]} - \frac{3}{2} \zeta^{mn} K_m^{[IJ} K_n^{KL]} + \frac{1}{12} \zeta^m  
K_{mn}^{[IJ} K^{n \, KL]}, \quad
Z^-_{IJKL} = \frac{1}{2} T^{mnp} \, K_{mn}^{[IJ}  K_{p}^{KL]},\notag \\[2mm]
F^{IJ} = \frac{1}{8}F^m K_m^{IJ} + \frac{1}{16}F^{mn} K_{mn}^{IJ}.\label{invertedid}
\end{gather}
Similarly, the background covariant derivative of the $\rm SO(7)$ tensors can be computed using the Killing spinor equation \eqref{killspineqn}
 \begin{gather}
  \mathring D_m \xi = 2 m_7 \xi_m,\quad
  \mathring D_m \xi_n = 6 m_7 \xi_{mn} - 2 m_7 \xi \mathring g_{mn},\quad
  \mathring D_p \xi_{mn} = \frac{1}{3} m_7 \left(\mathring g_{mn} \xi_p - \mathring g_{p(m} \xi_{n)} \right),\notag \\
  \mathring D_m \zeta = 2 m_7 \zeta_m,\quad
  \mathring D_m \zeta_n = 6 m_7 \zeta_{mn} - 2 m_7 \zeta \mathring g_{mn},\quad
  \mathring D_p \zeta_{mn} = \frac{1}{3} m_7 \left(\mathring g_{mn} \zeta_p - \mathring g_{p(m} \zeta_{n)} \right),\notag \\
  \mathring D_m S_{npq} = \frac{1}{6} m_7 \mathring \eta_{mnpqrst}  S^{rst},\qquad
  \mathring D_m T_{npq} = \frac{1}{6} m_7 \mathring \eta_{mnpqrst} T^{rst},\notag \\[2mm]
  \mathring D_n F_m = m_7 F_{mn},\qquad
  \mathring D_{p} F_{mn} = 2 m_7 \mathring g_{p[m} F_{n]}.
  \label{diffident}
\end{gather}
We stress once more that both  \eqref{invertedid} and \eqref{diffident} do not depend on the  particular forms of the $\rm SO(8)$ tensors  $Y^{\pm}_{IJKL}$, $Z^{\pm}_{IJKL}$ and $F_{IJ}$.

\subsection{Special identities}
\label{sec:so7iden}

The starting point  for proving  the identities satisfied by the  $\rm SO(7)$ tensors
and listed  in tables~\ref{so7id1}-\ref{so7id7}  are various contraction identities for the $\rm SO(8)$ tensors  $Y^{\pm}_{IJKL}$, $Z^{\pm}_{IJKL}$ and $F_{IJ}$. The  latter   follow directly from the definitions of these tensors in Eqs.\ \eqref{so8tensors} and \eqref{Ftens}, and can be split into  several groups depending on the number of factors and the  number of contractions. Each group then gives rise to different types of $\rm SO(7)$ identities. The identities given in this section are sufficient for determining the internal components of the metric and 3-form potential from the uplift ans\"atze and proving that the metric and 3-form potential thus obtained solve the field equations.   

\vspace{6 pt}

\noindent
\emph{\underline{A. Double contraction identities between two of the  $Y^\pm_{IJKL}$ and $Z^\pm_{IJKL}$ tensors}:}
\begin{align}\label{yzeqn1}
  Y^+_{IJMN}Y^+_{MNKL} & = Z^-_{IJMN}Z^-_{MNKL}, \qquad
  Y^-_{IJMN}Y^-_{MNKL} = Z^+_{IJMN}Z^+_{MNKL},  \\[2mm]
  \label{yzeqn5}
  Y^+_{IJMN}Z^+_{MNKL} & = Z^-_{IJMN}Y^-_{MNKL}, \qquad  Y^-_{IJMN}Z^-_{MNKL} = Z^+_{IJMN}Y^+_{MNKL},
\end{align}
and
\begin{align}
  Y^+_{IJMN}Y^-_{MNKL} &= Z^-_{IJMN}Z^+_{MNKL}, \qquad  Y^-_{IJMN}Y^+_{MNKL} = Z^+_{IJMN}Z^-_{MNKL}, \notag \\[2mm]
  Z^+_{IJMN}Y^-_{MNKL} &= Y^-_{IJMN}Z^+_{MNKL}, \qquad  Y^+_{IJMN}Z^-_{MNKL} = Z^-_{IJMN}Y^+_{MNKL}. \label{yzeqn3}
\end{align}
Note that each set of (anti-)selfdual tensors, $Y^{\pm}_{IJKL}$ and $Z^{\pm}_{IJKL}$, respectively, do not in themselves lead to simple quadratic identities, but are instead related to each other via quadratic relations. This is pertinent to the discussion in section~\ref{sec:harmonics} and appendix~\ref{app:singleset}, where it is argued that one can always make do with a single set of (anti-)selfdual tensors at the price of working to higher order. Here we see that there are no self-contained set of quadratic identities for a single set of (anti-)selfdual tensors. Therefore, the result is that one must work with expressions that are higher-order in tensors---as illustrated explicitly in appendix~\ref{app:singleset}. This is to be contrasted with the previously known uplifts where the situation is simpler, see table~\ref{table:comG2SU4-}. In the case of the G$_2$ invariant quantities, there are quadratic relations between the single set of (anti-)selfdual tensors. While in the slightly more complicated SU(4)$^-$ example, the 
single set of (anti-)selfdual 4-form tensors close on a 2-form tensor, rather than another set of 4-form tensors. More generally, for stationary points with even less symmetry the lesson seems to be that one must include enough (anti-)selfdual tensors in order to have quadratic relations between the tensors. Otherwise, the metric and 3-form potential will not be expressible at most quadratically in the SO(7) tensors.

\vspace{6 pt}

\noindent
\emph{\underline{B. Double contraction identities with triple factors}:}
\begin{equation}
\begin{split}\label{eqtr1}
 Y^+_{IJMN}Y^+_{MNPQ}Y^+_{PQKL}  & = 4Y^+_{IJKL},\qquad  Z^+_{IJMN}Z^+_{MNPQ}Z^+_{PQKL} = 4Z^+_{IJKL}, \\[2 mm]
\end{split}
\end{equation}
as well as
\begin{gather}
 Y^-_{IJMN}Y^-_{MNPQ}Y^-_{PQKL}  = 4Y^-_{IJKL},  \\[2 mm]
 Y^+_{IJMN}Y^-_{MNPQ}Y^+_{PQKL}  =0,\qquad    Y^-_{IJMN}Y^+_{MNPQ}Y^-_{PQKL}  =0,  \\[2 mm]
 Y^-_{IJMN}Y^+_{MNPQ}Y^+_{PQKL} + Y^+_{IJMN}Y^+_{MNPQ}Y^-_{PQKL}  = 4Y^-_{IJKL}, \\[2 mm]
 Y^+_{IJMN}Y^-_{MNPQ}Y^-_{PQKL} + Y^-_{IJMN}Y^-_{MNPQ}Y^+_{PQKL}  = 4Y^+_{IJKL},\label{eqtr2}
\end{gather}
and analogous identities obtained by replacing $Y$ by $Z$ in the above identities. 

\noindent
\emph{\underline{C. Identities involving the $F_{IJ}$ tensor}:}
\begin{gather}\label{yzeqn2}
  Y^+_{IKLM} Z^+_{JKLM} = Z^-_{IKLM} Y^-_{JKLM} = 12F_{IJ},  \\[2mm]
  \label{yzeqn61}
   Y^\pm_{IJKL} F^{KL} = Z^\pm_{IJKL} F^{KL} = 0,\\
\label{yzeqn62}  8 Y^\pm_{[IJK|M|} F^{M}{}_{L]} = \pm Z^\pm_{IJKL}, \quad
  8 Z^\pm_{[IJK|M|} F^{M}{}_{L]} = \mp Y^\pm_{IJKL}.
\end{gather}

Given the identities for the SO(8) tensors, it is clear from the inversion formulae \eqref{invertedid} that these identities imply identities satisfied by the SO(7) tensors in \eqref{ourtens}. We list these identities in tables \ref{so7id1}--\ref{so7id6}. Note that we do not use the cubic identities \eqref{eqtr2} in deriving the SO(7) tensor identities---they will be used in section~\ref{sec:gaugedsoln} to exponentiate the 56-bein in the unitary gauge.

While it is correct that the SO(7) tensor identities in tables \ref{so7id1}--\ref{so7id6} are a consequence of substituting the inversion formulae into the SO(8) tensor identities \eqref{yzeqn1}--\eqref{eqtr1} and \eqref{yzeqn2}, \eqref{yzeqn62}, it is rather laborious to obtain these identities by the said method---at least without the aid of a computer program. In appendix~\ref{app:idproofs}, we sketch a simpler proof for these identities. Furthermore, in the appendix we explain how the identities listed in tables \ref{so7id3}--\ref{so7id7} are derived from the identities in  tables \ref{so7id1}--\ref{so7id6}.  Despite the fact that the derivation of these identities is quite an involved task, we have tried to present the identities as systematically as possible. In particular, the order in which the identities are presented is such as to indicate the fact that identities listed prior to a given identity may have been used to derive or simplify that identity. This means that, for instance, we have included an identity that may be obtained by contracting another identity, allowing the reader to check the consistency of the two. In any case, here we limit the explanation of the derivations to the comments in the table captions, sketching a derivation of the identities in appendix~\ref{app:idproofs}.

\begin{table}[t]
\renewcommand{\arraystretch}{1.8}
\begin{center}
\caption{\label{so7id1}}
\scalebox{0.95}{
\begin{tabular}{@{\extracolsep{0 pt}} c c}
\toprule
(i) & $ \displaystyle \xi^{mn} \xi_{mn} = \frac{3}{2} + \frac{\xi^2}{6}, \quad \xi^p \xi_p
  = 9 - \xi^2, \quad
  \zeta^{mn} \zeta_{mn} = \frac{3}{2} + \frac{\zeta^2}{6},  \quad   \zeta^p \zeta_p = 9 - \zeta^2$ \\
(ii) &  $ \displaystyle  S_{mnp} S^{mnp} = 6, \quad T_{mnp}
  T^{mnp} = 6$ \\
(iii) & $ \xi^{mn} \xi_n = 0, \quad \zeta^{mn} \zeta_n = 0$ \\
(iv) & $ \displaystyle   \xi^{mp} {\xi}^{n}{}_{p} = \left( \frac{1}{4} - \frac{\xi^2}{36}
  \right) \mathring g^{mn} + \frac{\xi}{3} \xi^{mn} - \frac{1}{36}
  \xi^m \xi^n, \;\; \zeta^{mp} {\zeta}^{n}{}_{p} = \left( \frac{1}{4}
    - \frac{\zeta^2}{36} \right) \mathring g^{mn} + \frac{\zeta}{3}
  \zeta^{mn} - \frac{1}{36}
  \zeta^m \zeta^n $ \\[6 pt]
(v) & $ \displaystyle   S^{mpq} S^{n}{}_{pq} = \left( 1 -
    \frac{\zeta^2}{9} \right) \mathring g^{mn} - \frac{1}{9} \zeta^m
  \zeta^n + \frac{2\zeta}{3} \zeta^{mn}, \;\;
  T^{mpq} T^{n}{}_{pq} = \left( 1 - \frac{\xi^2}{9} \right) \mathring g^{mn} - \frac{1}{9} \xi^m \xi^n + \frac{2\xi}{3} \xi^{mn}$ \\[6 pt]
(vi)  & $ \displaystyle \mathring \eta_{mnqrstu} T^{qrs} T^{tu}{}_{p} =
  8 \xi_{[m} \xi_{n]p} - \frac{4}{3} \xi \xi _{[m} \mathring
  g_{n]p},\quad \mathring \eta_{mnqrstu} S^{qrs} S^{tu}{}_{p} = 8 \zeta_{[m} \zeta_{n]p} - \frac{4}{3} \zeta
  \zeta_{[m} \mathring g_{n]p}$ \\[6 pt]
(vii) & $ \displaystyle   S^{mnr} S_{pqr} = 2\zeta^{[m}{}_{[p}
  \zeta^{n]}{}_{q]} + \left(\frac{1}{2} - \frac{\zeta^2}{18}\right)
  \delta^{mn}_{pq} - \frac{1}{9} \zeta^{[m} \zeta_{[p} \delta^{n]}_{q]}$ \\[6 pt]
(viii)&  $ \displaystyle   T^{mnr} T_{pqr} = 2\xi^{[m}{}_{[p}
  \xi^{n]}{}_{q]} + \left(\frac{1}{2} - \frac{\xi^2}{18}\right)
  \delta^{mn}_{pq} - \frac{1}{9} \xi^{[m} \xi_{[p}
  \delta^{n]}_{q]}$ \\[6 pt]
\bottomrule
\end{tabular}
}
\\[6 pt]
Identities derived from \eqref{yzeqn1} and \eqref{eqtr1}.
\end{center}
\end{table}

\begin{table}[H]
\renewcommand{\arraystretch}{1.8}
\begin{center}
\caption{\label{so7id2}}
\begin{tabular}{@{\extracolsep{10 pt}} c c}
\toprule
 (i) & $\displaystyle \xi^m \zeta_m = - \xi \zeta, \qquad \xi^{mn} \zeta_{mn} =  \frac{1}{6} \xi \zeta, \qquad S_{mnp} T^{mnp} = 0 $ 
\\
(ii) & $\displaystyle  \mathring \eta^{mnpqrst} S_{npq} T_{rst} =  18 F^m$\\
(iii) & $\displaystyle\xi^{mn} \zeta_n = \frac{\xi}{6} \zeta^m - \frac{\zeta}{6} \xi^m  + \frac{3}{2} F^m, \quad
  \zeta^{mn} \xi_n = - \frac{\xi}{6} \zeta^m + \frac{\zeta}{6} \xi^m - \frac{3}{2} F^m$ \\
(iv) & $\displaystyle  \xi^{mp} {\zeta}^{n}{}_{p} = -\frac{1}{36} \xi \zeta \mathring g^{mn} - \frac{1}{36} \xi^m \zeta^n + \frac{1}{6}\left(\zeta \xi^{mn} + \xi \zeta^{mn}\right) + \frac{1}{4}  F^{mn}$\\
(v) & $\displaystyle  S^{m}{}_{pq}T^{npq} = -\frac{1}{9} \xi \zeta \mathring g^{mn} - \frac{1}{18} \left(\xi^m \zeta^n + \zeta^m  \xi^n\right) + \frac{1}{3} \left(\zeta \xi^{mn} + \xi \zeta^{mn}\right) - \frac{1}{2} F^{mn} $\\
(vi) & $\displaystyle\mathring \eta_{npqrstu} S_{m}{}^{qr} T^{stu} = -4 \xi_{m[n}\zeta_{p]} - 4 \zeta_{m[n}\xi_{p]}
  + \frac{2}{3} \zeta \mathring g_{m[n} \xi_{p]} + \frac{2}{3} \xi \mathring g_{m[n} \zeta_{p]} - 6 \mathring g_{m[n} F_{p]}$ \\
(vii) & $\displaystyle  \mathring \eta_{npqrstu} T_{m}{}^{qr} S^{stu} = -4 \xi_{m[n}\zeta_{p]} - 4 \zeta_{m[n}\xi_{p]}
  + \frac{2}{3} \zeta \mathring g_{m[n} \xi_{p]} + \frac{2}{3} \xi \mathring g_{m[n} \zeta_{p]} + 6 \mathring g_{m[n} F_{p]}$\\
(viii) &   $\displaystyle S^{mnr} T_{pqr} + T^{mnr} S_{pqr} = -\frac{1}{9} \xi \zeta \delta^{mn}_{pq} - \frac{1}{9}
  \xi^{[m} \zeta^{\phantom{[n}}_{[p} \delta^{n]}_{q]} - \frac{1}{9} \zeta^{[m}
  \xi^{\phantom{[n}}_{[p} \delta^{n]}_{q]} + 4 \xi^{[m}{}_{[p}
  \zeta^{n]}{}_{q]}$\\[6 pt]
\bottomrule
\end{tabular}
\\[6 pt]
Identities derived from \eqref{yzeqn5} and \eqref{yzeqn2}.
\end{center}
\end{table}

\begin{table}[ht]
\renewcommand{\arraystretch}{1.8}
\begin{center}
\caption{\label{so7id4}}
\begin{tabular}{@{\extracolsep{10 pt}} c c}
\toprule
(i) & $\displaystyle  S_{mnp} \xi^p + T_{mnp} \zeta^p = 0, \qquad  S_{mnp} \zeta^p = T_{mnp} \xi^p = 0$  \\
(ii) & $\displaystyle   S_{qmn} \zeta_{p}{}^q = S_{q[mn} \zeta_{p]}{}^{q} = \frac{\zeta}{3} S_{mnp} 
  - \frac{1}{36} \mathring \eta_{mnpqrst} \zeta^q S^{rst}$ \\
(iii) & $\displaystyle   T_{qmn} \xi_{p}{}^q = T_{q[mn} \xi_{p]}{}^{q} = \frac{\xi}{3} T_{mnp} 
  - \frac{1}{36} \mathring \eta_{mnpqrst} \xi^q T^{rst}$ \\ 
(iv) & $\displaystyle   4 \zeta^{qr}T_{rmn} - \frac{1}{9} \mathring \eta^{q}{}_{mnstuv} \zeta^s T^{tuv} = 8
  S^{sq}{}_{[m} \xi_{n]s} - \frac{4}{3}\xi S^q{}_{mn}$ \\ 
(v) & $\displaystyle   4 \xi^{qr}S_{rmn} - \frac{1}{9} \mathring \eta^{q}{}_{mnstuv} \xi^s S^{tuv} = 8
  T^{sq}{}_{[m} \zeta_{n]s} - \frac{4}{3}\zeta
  T^q{}_{mn} $\\[6pt]
\bottomrule
\end{tabular}
\\[6 pt]
Identities derived from \eqref{yzeqn3}.
\end{center}
\end{table}

\begin{table}[ht]
\renewcommand{\arraystretch}{1.8}
\begin{center}
\caption{\label{so7id6}}
\begin{tabular}{@{\extracolsep{10 pt}} c c}
\toprule
(i) & $\displaystyle  S^{mnp} F_{np} = 0, \quad 
  S^{mnp} F_p = \frac{1}{12} \mathring \eta^{mnpqrst} S_{pqr} F_{st}$\\
(ii) & $\displaystyle  
  T^{mnp} F_{np} = 0, \quad 
  T^{mnp} F_p = \frac{1}{12} \mathring \eta^{mnpqrst} T_{pqr} F_{st}$\\
(iii) & $\displaystyle  S^{q[mn} F^{p]}{}_q = \frac{2}{3}T^{mnp} + \frac{1}{18} \mathring \eta^{mnpqrst} S_{qrs} F_t$ \\
(iv) & $\displaystyle  T^{q[mn} F^{p]}{}_q = -\frac{2}{3}S^{mnp} + \frac{1}{18} \mathring  \eta^{mnpqrst} T_{qrs} F_t$\\[6pt]
\bottomrule
\end{tabular}
\\[6 pt]
Identities derived from \eqref{yzeqn61} and \eqref{yzeqn62}.
\end{center}
\end{table}

\begin{table}[H]
\renewcommand{\arraystretch}{1.6}
\begin{center}
\caption{\label{so7id3}}
\begin{tabular}{@{\extracolsep{10 pt}} c c}
\toprule
(i) & $\displaystyle  F_m \xi^m =\, \zeta, \quad  F_m \zeta^m =\, - \xi, \quad  
  F_{m} \xi^{mn} =\, \frac{1}{6} \zeta^n + \frac{\xi}{6} F^n, \quad
  F_{m} \zeta^{mn} =\, - \frac{1}{6} \xi^n +  \frac{\zeta}{6} F^n $ \\
(ii) & $\displaystyle  F_{mn} \xi^n =\, - \zeta_m - \xi F_m,  \quad
    F_{mp} \xi^{p}{}_{n} =\, \frac{\zeta}{6} \mathring g_{mn} - \frac{1}{6} F_m \xi_n - \zeta_{mn} +
  \frac{\xi}{6} F_{mn}$ \\ 
(iii) & $\displaystyle  F_{mn} \zeta^n =\, \xi_m - \zeta F_m, \quad
  F_{mp} \zeta^{p}{}_{n} =\, -\frac{\xi}{6} \mathring g_{mn} - \frac{1}{6} F_m \zeta_n + \xi_{mn} +
  \frac{\zeta}{6} F_{mn}$ \\
(iv) & $\displaystyle  F^m F_m =\, 1, \quad F^{mn} F_n = 0, \quad F^{mp} F_{pn} = F^m
  F_n - \delta^m_n $ \\[6pt] 
\bottomrule
\end{tabular}
\\[6 pt]
\parbox[c]{5 in} {$F$-tensor identities derived by contractions of the equations in
(iii) and (iv) in table~\ref{so7id2} with $\xi_m$, $\zeta_m$, $F_m$,
$\xi_{mq}$, $\zeta_{mq}$ and $F_{mq}$.} 
\end{center}
\end{table}

\begin{table}[ht]
\renewcommand{\arraystretch}{1.8}
\begin{center}
\caption{\label{so7id5}}
\begin{tabular}{@{\extracolsep{10 pt}} c c}
\toprule
(i) &  $\displaystyle \xi_{sm} S_{np}{}^s + \zeta_{sm} T_{np}{}^s = \xi_{s[m} S_{np]}{}^s + \zeta_{s[m} T_{np]}{}^s$ \\ 
(ii) &  $\displaystyle \xi_{s[m} S_{np]}{}^s = \frac{1}{9} (\zeta T_{mnp} + 2 \xi S_{mnp}) - \frac{1}{108}  \mathring\eta_{mnpqrst} (2 \zeta^q T^{rst} +  \xi^q S^{rst})$  \\ 
(iii) & $\displaystyle  \zeta_{s[m} T_{np]}{}^s = \frac{1}{9} (\xi S_{mnp} + 2 \zeta T_{mnp}) - \frac{1}{108}
  \mathring \eta_{mnpqrst} (2 \xi^q S^{rst} + \zeta^q
  T^{rst}) $\\[6pt] 
\bottomrule
\end{tabular}
\\[6 pt]
Identities derived from the equations in (iv) and (v) in table~\ref{so7id4}.
\end{center}
\end{table}

\begin{table}[ht]
\renewcommand{\arraystretch}{1.8}
\begin{center}
\caption{\label{so7id7}}
\begin{tabular}{@{\extracolsep{10 pt}} c c}
\toprule
(i) &  $\displaystyle \mathring \eta_{mnpqrst} \xi^p \zeta^q S^{rst} = 6 \zeta S_{mnp} \xi^p + 54 S_{mnp} F^{p},\quad
  \mathring \eta_{mnpqrst} \zeta^p \xi^q T^{rst} = 6 \xi T_{mnp} \zeta^p - 54 T_{mnp}  F^{p} $ \\
(ii) &  $\displaystyle \mathring \eta_{mnpqrst} F^p \zeta^q S^{rst} = 6S_{mnp} \xi^p + 6 \zeta S_{mnp}  F^{p}, \quad    \mathring \eta_{mnpqrst} F^p \xi^q T^{rst}=-6T_{mnp} \zeta^p + 6 \xi T_{mnp}  F^{p}
    $\\
(iii) & $ \displaystyle  \mathring \eta_{mnpqrst} F^p \xi^q S^{rst} = 6 \xi S_{mnp} F^p,\quad
  \mathring \eta_{mnpqrst} F^p \zeta^q T^{rst} = 6 \zeta T_{mnp} F^p$\\[6pt]
\bottomrule
\end{tabular}
\\[6 pt]
\parbox[c]{5.5 in}{Identities derived by contractions of the equations in (ii)--(iii) in
table~\ref{so7id4} with $\xi^p$, $\zeta^p$ and $F^p$; and contractions of (iii) and (iv) in table~\ref{so7id6} with $\xi_m$ and $\zeta_m$, respectively.}
\end{center}
\end{table}

\sect{The SO(3)$\times$SO(3) solution of gauged supergravity}
\label{sec:gaugedsoln}

In the unitary gauge defined in equation \eqref{coset}, the $u$ and $v$ matrices are of the form
\begin{equation}
  u_{IJ}{}^{KL} = \sum_{n=0}^\infty \frac{1}{(2n)!}  \left[ (\phi\phi^*)^n  \right]_{IJKL}, 
  \qquad v^{IJKL} = \sum_{n=0}^\infty \frac{1}{(2n+1)!} \left[ \phi^* (\phi\phi^*)^n \right]_{IJKL}.
\end{equation}

For an SO(3)$\times$SO(3) invariant configuration, the most general parametrisation of the scalar and pseudoscalar expectation value $\phi_{IJKL}$ is given by the SO(3)$\times$SO(3) invariant quantities defined in equation \eqref{so8tensors}
\begin{equation} \label{phi-general}
  \phi_{IJKL} = \frac{\lambda}{2} \left[ \cos \alpha \left( Y^+_{IJKL} + i Y^-_{IJKL} \right) 
  - \sin \alpha \left( Z^+_{IJKL} - i Z^-_{IJKL} \right) \right],
\end{equation}
where the parameter $\alpha$ may be freely chosen without loss of generality.  This is because, while the relevant SO(3)$\times$SO(3) invariant truncation of the theory contains two complex scalars, the potential corresponding to this truncation is invariant under an extra U(1) symmetry that lies outside the gauge group, namely SO(8) \cite{bobev11}.  The $\alpha$ parameter corresponds to this U(1) freedom that leaves the potential invariant.  In what follows we will choose to keep the value of $\alpha$ general.  Interestingly, from an eleven-dimensional perspective we find that $\alpha$ corresponds to a coordinate transformation of the eleven-dimensional solution along the seven compactified directions (see section~\ref{sec:alpha}).

In exponentiating the scalar expectation value $\phi_{IJKL}$ to find the $u$ and $v$ matrices, it is useful to define~\footnote{In what follows, we make use of the short-hand notation 
\begin{equation}
  A \, B = ( A \, B )_{IJKL} = A_{IJMN} B_{MNKL}.
\end{equation}}
\begin{equation}
  \label{Pi}
  \Pi = \frac{1}{8} \left( Y^+ + i Y^- \right) \left( Y^+ - i Y^- \right) = \frac{1}{8} \left( Z^+ -i Z^- \right) \left( Z^+ + i Z^- \right),
\end{equation}
which, using the cubic identities \eqref{eqtr1} and \eqref{eqtr2}, satisfies the following properties 
\begin{equation}
  \Pi^2 = \Pi, \qquad \Pi^*_{IJKL} = \Pi_{KLIJ}.
  \end{equation}
Therefore, $\Pi$ is a hermitean projector, and
\begin{equation}\label{}
\left( Y^+ - i Y^- \right) \Pi = Y^+ - i Y^-, \qquad \left( Z^+ + i Z^- \right)
  \Pi = Z^+ + i Z^-.
\end{equation}
In particular, using identities \eqref{eqtr1}, we find that
\begin{equation}
  \phi \phi^* = 2 \lambda^2 \Pi, \qquad \phi^* \Pi = \phi^*.
\end{equation}
Hence, the $u$ and $v$ matrices may be written as follows
\begin{gather}
  u_{IJ}{}^{KL} = \delta_{IJ}^{KL} + ( c - 1) \Pi_{IJKL}, \label{u} \\ \label{v}
  v^{IJKL} = \frac{s}{2\sqrt{2}} \left[ \cos \alpha (Y^+ - i Y^-) -
    \sin \alpha (Z^+ + i Z^-) \right]_{IJKL},
\end{gather}
where 
\begin{equation*}
 c = \cosh(\sqrt{2}\lambda),\qquad s = \sinh(\sqrt{2}\lambda).
\end{equation*}
The scalar potential for the scalar $\lambda$ reads
\begin{equation}\label{}
{\cal P} = - \frac{g^2}{2} (s^4-8s^2-12),
\end{equation}
and, indeed, does not depend on $\alpha$.

The SO(3)$\times$SO(3) invariant stationary point is given by 
\begin{equation}
 \frac{d \cal P}{d s} = 0\,,
\end{equation}
and corresponds to \cite{Warner84}
\begin{equation}
  c = \sqrt{5}, \qquad s = 2.
  \label{statval}
\end{equation}
This stationary point is the only known stable non-supersymmetric stationary point of $D=4$ maximal supergravity \cite{Fischbacher:2010ec,Fischbacherency}.  In fact, there clearly exists another stationary point corresponding to ${s \rightarrow -s}$, that is $s=-2$.  From the perspective of the $D=11$ solution this corresponds to ${A_{mnp} \rightarrow - A_{mnp}}$ under which the equations of motion \eqref{Einseqn1}-\eqref{maxwell} are  invariant. 
 We will take $s=2$ henceforth, while keeping this in mind.

\sect{The SO(3)$\times$SO(3) solution of $D=11$ supergravity}
\label{sec:11dsoln}

Given the scalar profile of the SO(3)$\times$SO(3) invariant solution of the gauged theory described in the previous section, the eleven-dimensional SO(3)$\times$SO(3) solution is simply constructed by applying the uplift formulae \eqref{ans:metric}  and \eqref{ans:flux} 
for the internal metric  and 3-form potential \cite{dWNW, dWN13}. In this section we present the details of the calculation leading to the solution in its simplified form.

\subsection{The internal metric}
\label{sec:metric}
We apply the uplift formula (\ref{ans:metric}) to evaluate the metric from the data 
at the  SO(3)$\times$SO(3) invariant stationary point.
The Sp(56) property of the $u$ and $v$ matrices \cite{dWNn8}
\begin{align}
 u^{MN}{}_{IJ} \, u_{MN}{}^{KL} - v_{MNIJ} \, v^{MNKL} &= \delta^{KL}_{IJ},\label{e7reln1} \\
 u^{MN}{}_{IJ} \, v_{MNKL} - v_{MNIJ} \, u^{MN}{}_{KL} &= 0, \label{e7reln2}
\end{align}
can be used to rewrite the scalar part of the metric ansatz \eqref{ans:metric} as follows
\begin{equation}
 \left( u^{MN}{}_{IJ} + v^{MNIJ} \right)\left( u_{MN}{}^{KL} + v_{MNKL} \right) = 
 -\delta_{IJ}^{KL} + 2 \text{Re} \left( u^{MN}{}_{IJ} u_{MN}{}^{KL} + v^{MNIJ} u_{MN}{}^{KL} \right).
\end{equation}
Substituting in the expressions for $u$ and $v$, equations \eqref{u} and \eqref{v}, we find that
\begin{equation}
 \text{Re} \left( u^{MN}{}_{IJ} u_{MN}{}^{KL} + v^{MNIJ} u_{MN}{}^{KL} \right) = 
 \delta_{IJ}^{KL} + s^2 \text{Re}(\Pi_{IJKL}) + \frac{sc}{2\sqrt{2}} \left( \cos \alpha \, Y^{+}- \sin \alpha \, Z^{+}
  \right)_{IJKL}.
\end{equation}
Contracting the expression above with $K^{m\,IJ} K^{n\,KL}$ and using the completeness relation \eqref{complete} to rewrite the expression in terms of SO(7) tensors gives
\begin{align}
  \Delta^{-1} g^{mn}(x,y) =&\, \mathring g^{mn} + \frac{s^2}{4} \left[
  \frac{1}{9} \mathring g^{m[n} \xi^{q]} \xi_q + 2 \xi^{mp}
  {\xi}^{n}{}_{p} + S^{mpq} S^n{}_{pq}
  + \frac{1}{9} \mathring g^{m[n} \zeta^{q]} \zeta_q + 2 \zeta^{mp}
  {\zeta}^{n}{}_{p} + T^{mpq} T^n{}_{pq} \right] \nonumber\\[2mm]
  &- \sqrt{2} s c \left( \cos \alpha \xi^{mn} - \sin \alpha \zeta^{mn}
  \right).
\end{align}
Using the SO(7) identities in table~\ref{so7id1}, the above expression reduces to
\begin{equation} \label{inmetricalpha}
\Delta^{-1} g^{mn} =\left[ c^2 - \frac{s^2}{18}(\xi^2 + \zeta^2) \right] \mathring
  g^{mn} - \frac{s^2}{18} (\zeta^m \zeta^n + \xi^m \xi^n) +
  \frac{s}{3}\left(\mathcal{X}_1 \xi^{mn} + \mathcal{Z}_1 \zeta^{mn} 
  \right)\,,
\end{equation}
where
\begin{align}
  \mathcal{X}_1(\alpha) = \xi s - 3\sqrt{2} c \cos\alpha , \qquad
  \mathcal{Z}_1(\alpha) = \zeta s + 3\sqrt{2} c \sin \alpha.
\end{align}

The first four lines of equations in tables \ref{so7id1} and \ref{so7id2} and the
identities in table~\ref{so7id3} can be used to invert the densitised  metric
(still for arbitrary $\alpha$)
\begin{equation} \label{metricalpha}
  \Delta g_{mn} = \frac{1}{\mathcal{X}_2^2 + 2 c^2 \mathcal{X}_2
    \mathcal{Z}_2 + \mathcal{Z}_2^2 + \mathcal{Y}} \Big[
  2(\mathcal{X}_1^2 + \mathcal{Z}_1^2) \mathring g_{mn} - 12 s
  (\mathcal{X}_1 \xi_{mn} + \mathcal{Z}_1 \zeta_{mn}) + s^2 f_m f_n  \Big],
\end{equation}
where
\begin{equation}\label{}
\begin{split}
 \mathcal{X}_2(\alpha) = \sqrt{2} \cos \alpha\, \xi s & - 3c\,, \qquad
 \mathcal{Z}_2(\alpha)   = -\sqrt{2} \sin \alpha\, \zeta s - 3c\,,\\[6 pt]  
  \mathcal{Y}(\alpha) & = s^4(\cos^2 \alpha - \sin^2 \alpha) (\xi^2 -
  \zeta^2),
\end{split}
\end{equation}
and 
\begin{equation}\label{}
f_m(\alpha) = \sqrt{2} c \cos \alpha\, \zeta_m + \sqrt{2} c \sin
  \alpha\, \xi_m + 3 s F_m\,.
\end{equation}

We can calculate the warp factor,  $\Delta$, using \eqref{warpbla}, by evaluating the variations 
\begin{equation}\label{vardg}
\Delta g_{mn}\,\delta(\Delta^{-1}g^{mn})  \,,
\end{equation}
with respect to $\alpha$ and $\lambda$. After simplifying \eqref{vardg} using identities in  tables~\ref{so7id1}, \ref{so7id2} and  \ref{so7id3}, one can integrate back to obtain $\Delta$, with the overall normalisation fixed by requiring that $\Delta=1$ for $\lambda=0$. This gives
\begin{equation}\label{thewarpD}
\Delta^3\eql \frac{36}{\mathcal{X}_2^2 + 2 c^2 \mathcal{X}_2
    \mathcal{Z}_2 + \mathcal{Z}_2^2 + \mathcal{Y}}\,.
\end{equation}
This completes the derivation of the uplifted metric tensor, $g_{mn}$, for arbitrary values of $\lambda$ and $\alpha$.

\subsection{The internal flux}
\label{sec:flux}

As before, we simplify the scalar part of the flux ansatz \eqref{ans:flux} using the Sp(56) property of the $u$ and $v$ matrices
\begin{equation}
 \left( u^{ij}{}_{IJ} - v^{ij IJ} \right) \left( u_{ij}{}^{KL} + v_{ij KL} \right)
= \delta^{IJ}_{KL} + 2 i \, \textup{Im} \big(u^{ij}{}_{IJ} u_{ij}{}^{KL} - v^{ij IJ} u_{ij}{}^{KL} \big).
\end{equation}
For the $u$ and $v$ matrices corresponding to the SO(3)$\times$SO(3) invariant sector
\begin{equation}
 \textup{Im} \big( u^{ij}{}_{IJ} u_{ij}{}^{KL} - v^{ij IJ} u_{ij}{}^{KL} \big) = 
 s^2\, \textup{Im} (\Pi_{IJKL}) + \frac{sc}{2\sqrt{2}} Y^-_{IJKL}.
\end{equation}
Contracting the above expression with $K_{mn}^{IJ} K^{q\, KL}$ and making use of the completeness relation \eqref{complete}, the flux ansatz \eqref{ans:flux} gives
\begin{equation}\label{}
\begin{split}
 \Delta^{-1} g^{pq} A_{mnp}(x,y) =& \frac{s^2}{48\sqrt{2}} \bigg( 8
  S^{sq}{}_{[m} \xi_{n]s} - \frac{4}{3}\xi
  S^q{}_{mn} - 4 \xi^{qr}S_{rmn} +
  \frac{1}{9} \mathring \eta^{q}{}_{mnstuv} \xi^s
  S^{tuv} \\[6 pt] & \qquad\quad  - 8 T^{sq}{}_{[m} \zeta_{n]s} 
   + \frac{4}{3}\zeta T^q{}_{mn}+ 4 \zeta^{qr}T_{rmn} - \frac{1}{9} \mathring \eta^q{}_{mnstuv} \zeta^s T^{tuv} \bigg) \\[6 pt] & +
  \frac{1}{6} s c \left( \cos \alpha \,
    S^q{}_{mn} + \sin \alpha \, T^q{}_{mn}
  \right).
\end{split}
\end{equation}
Upon use of the identities in tables \ref{so7id4} and \ref{so7id5},
the expression above simplifies significantly:
\begin{equation}
  \Delta^{-1} g^{pq} A_{mnp} = \frac{s}{6\sqrt{2}} \left( 2 s S^{sq}{}_{[m} \xi_{n]s} - 2 s T^{sq}{}_{[m}
  \zeta_{n]s} - \frac{1}{3} \mathcal{X}_1 S^q{}_{mn} + \frac{1}{3} \mathcal{Z}_1 T^q{}_{mn} \right).
\label{Audd}
  \end{equation}
Multiplying the above equation by the metric and substituting the expression \eqref{metricalpha} for $\Delta g_{pq}$, and making full and repeated use of the SO(7) identities in section~\ref{sec:so7iden}, the resulting expression reduces to
\begin{equation} \label{Aalpha}
\begin{split} 
   A_{mnp} =  \frac{\Delta^3}{18\sqrt 2}\,  \bigg[
   & - \Big( \frac{s}{2}(1+c^2) \mathcal{X}_1 + \frac{s^3c}{6}\sqrt{2}\, \zeta \left( \sin \alpha\, \xi + \cos \alpha \,\zeta \right) \Big) S_{mnp}
   \\[6 pt] &   + \Big( \frac{s}{2}(1+c^2) \mathcal{Z}_1 - \frac{s^3c}{6}\sqrt{2}\, \xi \left( \sin  \alpha \,\xi + \cos \alpha\, \zeta \right) \Big) T_{mnp}
\\[6 pt] &  + \frac{s^2}{108}
  \mathring \eta_{mnpqrst} \left( \mathcal{Z}_1 \xi^q - \mathcal{X}_1
    \zeta^q \right) 
   \left( \mathcal{Z}_1 S^{rst} + \mathcal{X}_1 T^{rst} \right)
   \\[6 pt] &  -\frac{s^3}{6} \left( \mathcal{X}_1 \xi_{[m} + \mathcal{Z}_1
    \zeta_{[m}\right) S_{np]q} \xi^q
   \,   \bigg]\,.
\end{split}
\end{equation}
with $\Delta$ given in \eqref{thewarpD}.

Note that while it is clear that the metric  obtained from the ansatz \eqref{ans:metric} is manifestly symmetric in its indices, this is not the case for the 3-form potential \eqref{ans:flux}. However, as is shown in Ref.~\cite{dWN13}, the antisymmetry property of the 3-form potential is guaranteed to hold even off-shell for any values of the scalar fields as is the case for the 3-form potential in \eqref{Aalpha}.

This concludes the uplift of the SO(3)$\times$SO(3) stationary point to $D=11$ supergravity. It is indeed remarkable that such a complicated solution as this one can be so simply derived in the matter of a few calculational steps.

\subsection{Choice of $\alpha$}
\label{sec:alpha}

As remarked earlier, from the point of view of gauged supergravity we
are free to choose $\alpha$ without loss of generality, because of an
accidental U(1) symmetry of the potential that is outside the gauge group.  This is a novel feature of the SO(3)$\times$SO(3) invariant truncation and is absent for other truncations for which the higher dimensional uplift is known.  There ought to be a way of understanding this redundancy in the choice of $\alpha$ from an eleven-dimensional perspective.  Given that in the four-dimensional theory the U(1) transformation does not lead to a different stationary point, it must be the case that for any choice of $\alpha$ the uplifted solutions are equivalent, \textit{viz.}\ they are related by coordinate transformations as we demonstrate here. Specifically, we find that a shift in the parameter $\alpha$ corresponds to a diffeomorphism in the seven compactified 
dimensions, in the sense that 
\begin{equation}
  \delta_\alpha \left( \Delta g_{mn}(\alpha) \right) = \mathcal{L}_V \left( \Delta g_{mn}(\alpha) \right), \qquad
  \delta_\alpha \left( A_{mnp}(\alpha) \right) = \mathcal{L}_V \left( A_{mnp}(\alpha) \right),
\end{equation}
with the generating vector field $V$~\footnote{Note that, while $V$ is a Killing vector on the background internal space, corresponding to the round $S^7$, it is no longer a Killing vector in the deformed space given by the metric $g_{mn}$. In deforming the round seven-sphere to obtain the SO(3)$\times$SO(3) invariant solution, the number of Killing vector fields reduces from 28 to 6; these are given by $K^{12}$, $K^{13}$, $K^{23}$, $K^{67}$, $K^{68}$ and $K^{78}$.}
\begin{equation}
 V = -\frac{1}{2m_7} F^m \mathring D_m.
\end{equation}
This allows us to pick any particular value of $\alpha$: checking the equations of motion for that
particular value then implies that the equations are also satisfied for other values of $\alpha$.
Henceforth, we choose to fix the value of $\alpha$,
\begin{equation}
 \alpha = -\frac{\pi}{4},
\end{equation}
so that the metric \eqref{metricalpha} is symmetric under the interchange of tensors defined with respect to $Y^{\pm}_{IJKL}$ and $Z^{\pm}_{IJKL}.$ 
In this case,
\begin{gather}
  \sin (\alpha) =\, -\frac{1}{\sqrt{2}}, \qquad \cos (\alpha) = \, \frac{1}{\sqrt{2}}, \qquad \mathcal{Y} =\, 0, \\[2mm]
  \mathcal{X}_1 =\, \mathcal{X}_2 \equiv  \mathcal{X} = \xi s - 3 c, \qquad 
  \mathcal{Z}_1 =\, \mathcal{Z}_2 \equiv  \mathcal{Z} = \zeta s - 3 c,
\end{gather}
and the metric determinant is:
\begin{equation}\label{Delta}
  \Delta = 36^{1/3} \left( \mathcal{X}^2 + 2 c^2 \mathcal{X} \mathcal{Z} + \mathcal{Z}^2 \right)^{-1/3}.
\end{equation}

In summary, at the stationary point values given by equation \eqref{statval}, we find the internal metric and 3-form potential given in equations \eqref{soln:met} and \eqref{soln:3form}. 
It is only at the stationary point values, given in equation \eqref{statval}, that these expressions solve the equations of motion \eqref{Einseqn1}--\eqref{maxwell}. 
Note also that with the choice of $\alpha$ given in this section, the
metric is indeed symmetric under the interchange of tensors defined using
invariants $Y^{\pm}_{IJKL}$ and $Z^{\pm}_{IJKL},$ while the 3-form is antisymmetric.~\footnote{Note that under this interchange we also have $$F_{m} \rightarrow -F_{m}, \qquad F_{mn} \rightarrow -F_{mn}.$$ We refer the reader to the first equation in table~\ref{so7id4} for the antisymmetry of the last term in equation (\ref{soln:3form}). \label{Finter}} Given the symmetric form of the solution for the choice of $\alpha = -\pi/4$, this is the solution that we work with in order to verify that the field equations are satisfied.
 
\sect{Verification of the Einstein and Maxwell equations}
\label{sec:verif-einst-equat}

In this section, we verify that the SO(3)$\times$SO(3) invariant solution does indeed satisfy the field equations of $D=11$ supergravity, equations \eqref{Einseqn1}--\eqref{maxwell}. It is a surprising fact that the verification forms by far the most involved part of the work and requires the use of many of the identities listed in section~\ref{sec:so7iden}. In comparison, finding the solution using the non-linear ans\"atze is fairly straightforward. This is a testimony to the power of the uplift ans\"atze, which are non-linear. From the perspective of the SU(8) invariant reformulation, it is clear that the ans\"atze should lead to internal metric and 3-form potential components that satisfy the $D=11$ supergravity equations of motion. This is because they have been derived by the use of supersymmetry transformations which are first order equations, rather than second order as in the case of the field equations. Moreover, the highly non-linear 
problem of relating the scalars of the $D=4$ maximal gauged supergravity to the components of the internal metric and 3-form has been linearised by packaging the components of the $D=11$ fields in the generalised vielbeine. The relation between the scalars of the $D=4$ theory and the generalised vielbeine is a linear one. Both of the simplifications alluded to above mean that while the derivation of the solution is relatively simple, its verification in the context of the original formulation of $D=11$ supergravity \cite{CJS} becomes non-trivial.

In order to verify the Einstein and Maxwell equations \eqref{Einseqn1}--\eqref{maxwell}, we make use of the computer algebraic manipulation program \texttt{FORM} \cite{form} to simplify the expressions for the Ricci tensor and the 4-form field strength.

\subsection{Components of the Ricci tensor}
\label{sec:ricci-tensor}

We begin by computing the components of the eleven-dimensional Ricci tensors $R_{\mu}{}^{\nu}$ and $R_{m}{}^{n}$ that appear in the equations of motion, \eqref{Einseqn1} and \eqref{Einseqn2}, and whose indices are raised with the full metric, $g_{MN}$.  Denoting
\begin{equation}
 g^{\mu\nu}(x,y) = \Delta(y) \mathring g^{\mu\nu}(x), \qquad g_{\mu\nu}(x,y) = \Delta^{-1}(y) \mathring g_{\mu\nu}(x),
\end{equation}
the Christoffel symbols with mixed index components are
\begin{gather}
  \Gamma_{mn}^\rho = \Gamma_{m\nu}^p = 0, \\
  \Gamma_{\mu\nu}^p = -\frac{1}{2} g^{pq} \partial_q g_{\mu\nu} = \frac{1}{2}  \left(\Delta^{-1} \mathring D_q \Delta \right) g^{pq} g_{\mu\nu},\\
  \Gamma_{\mu n}^\rho = \frac{1}{2} g^{\rho\sigma} \partial_n g_{\mu\sigma} = -\frac{1}{2}\left(\Delta^{-1}
  \mathring D_n \Delta\right) \delta_\mu^\rho.
\end{gather}
Moreover, for convenience, we define
\begin{align}
  \hat\Gamma_{mn}^p = \Gamma_{mn}^p - \mathring \Gamma_{mn}^p = \frac{1}{2}g^{pq} \left( \mathring D_m g_{nq} + \mathring D_n g_{mq}
    - \mathring D_q g_{mn} \right).
\end{align}

The relevant components of the eleven-dimensional Riemann tensor are
 \begin{align}
  R^\mu{}_{\nu\rho\sigma} =& - \partial_\rho \Gamma_{\sigma\nu}^\mu + \partial_\sigma \Gamma_{\rho\nu}^\mu -
  \Gamma_{\rho M}^\mu \Gamma_{\sigma\nu}^M + \Gamma_{\sigma N}^\mu
  \Gamma_{\rho\nu}^N \
  =\ \mathring R^\mu{}_{\nu\rho\sigma} - \Gamma_{\rho m}^\mu  \Gamma_{\sigma\nu}^m + \Gamma_{\sigma n}^\mu \Gamma_{\rho\nu}^n  \nonumber\\[3pt]
  =& \mathring R^\mu{}_{\nu\rho\sigma} + \frac{1}{2} (\Delta^{-1} \mathring D_p \Delta ) (\Delta^{-1}
    \mathring D_q \Delta) g^{pq}  \delta_{[\rho}^\mu g^{\phantom{\mu}}_{\sigma]\nu},\\[3pt]
   R^{\mu}{}_{m \nu n} =& - \partial_\nu \Gamma_{nm}^\mu  + \partial_n \Gamma_{\nu m}^\mu - \Gamma_{\nu p}^\mu \Gamma_{nm}^p +
  \Gamma_{n \rho}^\mu \Gamma_{\nu m}^\rho \nonumber\\[3pt]
  =& -\frac{1}{2} \mathring D_n ( \Delta^{-1} \mathring D_m \Delta )\delta_\nu^\mu + \frac{1}{2}\hat\Gamma_{mn}^p
  (\Delta^{-1} \mathring D_p \Delta ) \delta_\nu^\mu + \frac{1}{4} ( \Delta^{-1} \mathring D_n \Delta ) ( \Delta^{-1} \mathring D_m \Delta) \delta_\nu^\mu,\\[2mm]
  R^m{}_{\mu n \nu} =& g^{mp} g_{\mu\rho} R^\rho{}_{p \nu n},\\[2mm]
 R^m{}_{npq} =& \mathring R^m{}_{npq} - \mathring D_p
  \hat\Gamma_{qn}^m + \mathring D_q \hat\Gamma_{pn}^m - \hat\Gamma_{p
    r}^m \hat\Gamma_{qn}^r + \hat\Gamma_{q r}^m \hat\Gamma_{pn}^r,
\end{align}
where $\mathring R^\mu{}_{\nu\rho\sigma}$ and $\mathring R^m{}_{npq}$ denote the Riemann tensors of the background $AdS_4$ and round seven-sphere, respectively.  The associated Ricci tensors in our conventions are given in \eqref{ricciads}.

It is now straightforward to obtain the expressions for the relevant components of the Ricci tensor,
\begin{align}
  \label{eq:105}
  R_{\mu\nu} &= R^\rho{}_{\mu\rho\nu} + R^p{}_{\mu p \nu} \nonumber\\[3pt]
  &= 3 \Delta m_4^2 g_{\mu\nu} + g_{\mu\nu} g^{mn} \left( (\Delta^{-1} \mathring D_m \Delta)
 (\Delta^{-1} \mathring D_n \Delta) -\frac{1}{2} \mathring D_n ( \Delta^{-1} \mathring D_m \Delta ) +
    \frac{1}{2}\hat\Gamma_{mn}^p ( \Delta^{-1} \mathring D_p \Delta ) \right), \\
  R_{mn} &= R^p{}_{mpn} + R^\rho{}_{m \rho n} \nonumber\\[3pt]
  &= -6 m_7^2 \mathring g_{mn} - \mathring D_p \hat\Gamma_{nm}^p +  \mathring D_n \hat\Gamma_{pm}^p - \hat\Gamma_{p r}^p
  \hat\Gamma_{nm}^r + \hat\Gamma_{n r}^p \hat\Gamma_{pm}^r \nonumber\\[3pt]
  &\quad +  (\Delta^{-1} \mathring D_m \Delta ) (\Delta^{-1} \mathring D_n \Delta )
  - 2 \mathring D_n ( \Delta^{-1} \mathring D_m \Delta )  + 2 \hat\Gamma_{mn}^p ( \Delta^{-1} \mathring D_p \Delta ).
\end{align}
In fact, it is more convenient for us to directly calculate $\Delta^{-1}R_\mu{}^\nu = \Delta^{-1}R_{\mu\rho}g^{\rho\nu}$
and $\Delta^{-1}R_m{}^n = R_{mp} (\Delta^{-1}g^{pn}$).  Using the expression for the internal metric given in equation \eqref{soln:met} and the expression for the determinant \eqref{det} as well as equations \eqref{diffident} and the SO(7) identities in section~\ref{sec:so7iden},
\begin{align}
  \Delta^{-1} R_\mu{}^\nu &= 3 m_4^2 \delta_\mu^\nu + \frac{ m_7^2 \, \Delta^{6}}{972} \bigg( 91 \mathcal{X}^4 - 140 \mathcal{X}^3\mathcal{Z} -
  718 \mathcal{X}^2\mathcal{Z}^2 - 140 \mathcal{X}\mathcal{Z}^3 + 91 \mathcal{Z}^4  \nonumber\\
  & \quad+ 24\sqrt{5}(\mathcal{X}+\mathcal{Z})(19\mathcal{X}^2 - 50\mathcal{X}\mathcal{Z} + 19\mathcal{Z}^2) + 1260(5\mathcal{X}^2
  +2\mathcal{X}\mathcal{Z} + 5\mathcal{Z}^2) \bigg) \, \delta_\mu^\nu,
  \label{ricci4d}
\end{align}
\begin{align}
  \Delta^{-1} R_m{}^n &= \frac{m_7^2 \, \Delta^{6}}{1296} \, \big(
  A_0(\mathcal{X},\mathcal{Z}) \delta_m^n + A_1(\mathcal{X},\mathcal{Z})\xi_m{}^n + A_1(\mathcal{Z},\mathcal{X})\zeta_m{}^n +
  A_2(\mathcal{X},\mathcal{Z}) F_m{}^n \nonumber\\[3pt]
  &\quad + A_3(\mathcal{X},\mathcal{Z}) \xi_m \xi^n
  + A_3(\mathcal{Z},\mathcal{X}) \zeta_m \zeta^n +
  A_4(\mathcal{X},\mathcal{Z}) {F}_m F^n +
  A_5(\mathcal{X},\mathcal{Z}) \xi_m \zeta^n +
  A_5(\mathcal{Z},\mathcal{X}) \zeta_m \xi^n \nonumber\\[3pt]
  &\quad +A_6(\mathcal{X},\mathcal{Z}) \xi_m F^n
  - A_6(\mathcal{Z},\mathcal{X}) \zeta_m F^n +
  A_7(\mathcal{X},\mathcal{Z}) {F}_m \xi^n -
  A_7(\mathcal{Z},\mathcal{X}) {F}_m \zeta^n \big).
  \label{riccimn}
\end{align}
Recall that in our conventions, the index $n$ on the left hand side is raised with the inverse metric $g^{mn}$, while on the right hand side we use the inverse metric on the round $S^7$, $\go^{mn}.$ 
The coefficient functions in the above equation are as follows:
\begin{align*}
  A_0(\mathcal{X},\mathcal{Z}) &= \frac{2\sqrt{5}}{3}(\mathcal{X} + \mathcal{Z}) \left(17\mathcal{X}^4 - 80\mathcal{X}^3\mathcal{Z} -
    66\mathcal{X}^2\mathcal{Z}^2 - 80\mathcal{X}\mathcal{Z}^3 + 17\mathcal{Z}^4\right) \nonumber\\
    &\quad + \frac{40}{3}\big(13\mathcal{X}^4 - 134\mathcal{X}^3\mathcal{Z} - 214\mathcal{X}^2\mathcal{Z}^2
  - 134\mathcal{X}\mathcal{Z}^3 + 13\mathcal{Z}^4\big)  \nonumber\\[2pt]
  &\quad +  40\sqrt{5}(\mathcal{X}+\mathcal{Z}) \left(17\mathcal{X}^2 -
    58\mathcal{X}\mathcal{Z} + 17\mathcal{Z}^2\right) -
  840(5\mathcal{X}^2 +2\mathcal{X}\mathcal{Z} + 5\mathcal{Z}^2),\\[4pt]
  A_1(\mathcal{X},\mathcal{Z}) &= -10080\sqrt{5}(\mathcal{X}^2 - \mathcal{Z}^2) - 96\left( 41 \mathcal{X}^3 - 45 \mathcal{X}^2
    \mathcal{Z} - 9\mathcal{X} \mathcal{Z}^2 - 35 \mathcal{Z}^3 \right) \nonumber\\[3pt]
 &\quad - 8\sqrt{5} (\mathcal{X} + \mathcal{Z}) \big(  17\mathcal{X}^3 - 55 \mathcal{X}^2 \mathcal{Z} - 33\mathcal{X} \mathcal{Z}^2 - 25 \mathcal{Z}^3 \big),\\[2mm]
  A_2(\mathcal{X},\mathcal{Z}) &=  10080(\mathcal{X}^2 - \mathcal{Z}^2)  + 96\sqrt{5}(\mathcal{X} - \mathcal{Z})\left( 7 \mathcal{X}^2 + 10
    \mathcal{X} \mathcal{Z} + 7\mathcal{Z}^2 \right) + 200 (\mathcal{X} - \mathcal{Z})(\mathcal{X} +  \mathcal{Z})^3,\\[2mm]
  A_3(\mathcal{X},\mathcal{Z}) &=  672\sqrt{5} (5\mathcal{X} + 13 \mathcal{Z}) + 32(45\mathcal{X}^2 - 160\mathcal{X}\mathcal{Z} +
  79\mathcal{Z}^2) \\[2pt]
  &\quad + \frac{8\sqrt{5}}{3} \left( 17\mathcal{X}^3 + 43 \mathcal{X}^2 \mathcal{Z} - 149\mathcal{X} \mathcal{Z}^2 + 17 \mathcal{Z}^3 \right),
\end{align*}
\begin{align*}
  A_4(\mathcal{X},\mathcal{Z}) &= -2016(\mathcal{X}^2 + 10\mathcal{X}\mathcal{Z} + \mathcal{Z}^2) - 96\sqrt{5} (\mathcal{X}
  + \mathcal{Z})(\mathcal{X}^2 - 50\mathcal{X}\mathcal{Z} + \mathcal{Z}^2),\\[2mm]
  A_5(\mathcal{X},\mathcal{Z}) &= -672\sqrt{5} (13\mathcal{X} + 5  \mathcal{Z}) - 64(50\mathcal{X}^2 - 33\mathcal{X}\mathcal{Z} -
  5\mathcal{Z}^2) \\
  &\quad - \frac{8\sqrt{5}}{3} \left( 81\mathcal{X}^3 - 61 \mathcal{X}^2 \mathcal{Z} + 75\mathcal{X} \mathcal{Z}^2 + 25 \mathcal{Z}^3 \right),\\[2mm]
  A_6(\mathcal{X},\mathcal{Z}) &= 336\sqrt{5}(\mathcal{X}^2 + 10\mathcal{X}\mathcal{Z} + \mathcal{Z}^2) + 16\left( 5\mathcal{X}^3
    - 188 \mathcal{X}^2 \mathcal{Z} - 175\mathcal{X} \mathcal{Z}^2 - 38 \mathcal{Z}^3 \right), \\[2mm]
  A_7(\mathcal{X},\mathcal{Z}) &= - 4032 (5\mathcal{X} + 13\mathcal{Z}) - 48\sqrt{5}(35\mathcal{X}^2 -
  118\mathcal{X}\mathcal{Z} + 47\mathcal{Z}^2) \\[3pt]
  &\quad - 16\left( 25\mathcal{X}^3 + 116\mathcal{X}^2 \mathcal{Z} - 75\mathcal{X} \mathcal{Z}^2 + 66 \mathcal{Z}^3 \right).
\end{align*}

Note that, like the metric, both $R_{\mu}{}^{\nu}$ and $R_{m}{}^{n}$ are symmetric under the interchange of tensors defined using $Y^{\pm}_{IJKL}$ and $Z^{\pm}_{IJKL}$, definitions \eqref{so7tensors}.~\footnote{See footnote \ref{Finter}.}

\subsection{4-form field strength}
\label{sec:4-form-field}

In this section, we calculate the 4-form field strength
\begin{equation}
  F_{mnpq} = 4! \mathring D_{[m} A_{npq]}
\end{equation}
of the 3-form potential given in equation \eqref{soln:3form}.  Using the equations for the derivatives of the SO(7) tensors \eqref{diffident}
\begin{align}
  F_{mnpq}& =  \frac{2 \sqrt{2}\, m_7 \, \Delta^{6}}{81} \bigg[
  B_1(\mathcal{X},\mathcal{Z}) \xi_{[m} S_{npq]} - B_1(\mathcal{Z},\mathcal{X}) \zeta_{[m} T_{npq]}
  + B_2(\mathcal{X},\mathcal{Z}) \zeta_{[m} S_{npq]} \notag \\[2mm]
   & \qquad \qquad -  B_2(\mathcal{Z},\mathcal{X}) \xi_{[m} T_{npq]} + B_3(\mathcal{X},\mathcal{Z}) \mathring \eta_{mnpqrst} S^{rst} - B_3(\mathcal{Z},\mathcal{X}) \mathring \eta_{mnpqrst} T^{rst}
\notag \\[2mm]
&\qquad \qquad \qquad \qquad + \mathring \eta_{mnpqrst} S^{rsu} \xi_u \left( B_4(\mathcal{X},\mathcal{Z}) \xi^t + B_4(\mathcal{Z},\mathcal{X}) \zeta^t
+  B_5(\mathcal{X},\mathcal{Z}) F^t \right) \bigg],
\label{F}
\end{align}
where we have simplified some expressions using the SO(7) identities in section~\ref{sec:so7iden} and
\begin{align*}
  B_1(\mathcal{X},\mathcal{Z}) &=  3(\mathcal{X}^2 + 10\mathcal{X}\mathcal{Z}
  + 49\mathcal{Z}^2)- \sqrt{5}(\mathcal{X}^2-\mathcal{Z}^2)(\mathcal{X} + 11\mathcal{Z}), \\[2mm]
  B_2(\mathcal{X},\mathcal{Z}) &=  3(5\mathcal{X}^2 + 2\mathcal{X}\mathcal{Z} + 5\mathcal{Z}^2) + \sqrt{5}(\mathcal{X}^3
  - 3\mathcal{X}^2\mathcal{Z} - 21 \mathcal{X} \mathcal{Z}^2 - \mathcal{Z}^3), \\
  B_3(\mathcal{X},\mathcal{Z}) &= -\frac{1}{8} (\mathcal{X} + 2 \mathcal{Z})(\mathcal{X}^2 - 2 \mathcal{X} \mathcal{Z}
  + 5 \mathcal{Z}^2) + \frac{\sqrt{5}}{48} (\mathcal{X}^2 + \mathcal{Z}^2)(\mathcal{X}^2 + 10 \mathcal{X} \mathcal{Z} + \mathcal{Z}^2), \\
  B_4(\mathcal{X},\mathcal{Z}) &= -\frac{5\sqrt{5}}{2} (\mathcal{X}^2 - \mathcal{Z}^2),\\[2mm]
  B_5(\mathcal{X},\mathcal{Z}) &=  15(\mathcal{X}^2 - \mathcal{Z}^2).
\end{align*}

Raising the indices on $F_{mnpq}$ using the inverse metric $g^{mn}$ poses the greatest challenge from a computational point of view. Therefore, we choose to calculate it using the following method
\begin{align}
  \Delta^{-1} F^{mnpq} 
  &=\, 4! \Delta^{3} (\Delta^{-1} g^{r[m}) (\Delta^{-1} g^{n|s|} )
  \bigg[ \Delta^{-1} g^{p|t|} \mathring D_r \Big( \Delta^{-1}
    A^{q]}{}_{st} \Big) - \mathring D_r \Big( \Delta^{-1} g^{p|t|}
  \Big) \Delta^{-1} A^{q]}{}_{st} \bigg].
\end{align}
Substituting the expression for the inverse metric, equation \eqref{inmetricalpha} and flux, equation \eqref{Audd} at the stationary point values and with $\alpha = - \pi/4$, and simplifying the resulting expression using equations \eqref{diffident} and the SO(7) identities in section~\ref{sec:so7iden} gives
\begin{align}
  \Delta^{-1}& F^{mnpq} = \frac{\sqrt{2}\, m_7 \, \Delta^{3}}{36}\bigg[
  C_1(\mathcal{X},\mathcal{Z}) \xi^{[m} S^{npq]}   - C_1(\mathcal{Z},\mathcal{X}) \zeta^{[m} T^{npq]}
  + C_2(\mathcal{X},\mathcal{Z}) \zeta^{[m} S^{npq]} \nonumber\\[2mm]
  &\hspace{27mm} - C_2(\mathcal{Z},\mathcal{X}) \xi^{[m} T^{npq]} + C_3(\mathcal{X},\mathcal{Z}) \mathring \eta^{mnpqrst} S_{rst}
  - C_3(\mathcal{Z},\mathcal{X}) \mathring \eta^{mnpqrst} T_{rst} \notag \\[2mm]
  &\hspace{40mm} + \mathring \eta^{mnpqrst} S_{rsu} \xi^u \left( C_4(\mathcal{X},\mathcal{Z}) \xi_t + C_4(\mathcal{Z},\mathcal{X}) \zeta_t +
    C_5(\mathcal{X},\mathcal{Z}) F_t \right) \bigg],
    \label{Finv}
\end{align}
where
\begin{align*}
  C_1(\mathcal{X},\mathcal{Z}) &= \frac{224}{3}\mathcal{Z}^2 - \frac{8}{9}(3 + \sqrt{5}\mathcal{Z}) \left( \mathcal{X}^2 +
    10\mathcal{X}\mathcal{Z} + \mathcal{Z}^2 \right),\\
  C_2(\mathcal{X},\mathcal{Z}) &= -\frac{224}{3}\mathcal{X}\mathcal{Z} + \frac{8}{9}(15 + \sqrt{5}\mathcal{X}) \left( \mathcal{X}^2 +
    10\mathcal{X}\mathcal{Z} + \mathcal{Z}^2 \right),\\
  C_3(\mathcal{X},\mathcal{Z}) &= -14(\mathcal{X} + 5\mathcal{Z}) - \frac{\sqrt{5}}{3} \left( 3\mathcal{X}^2 + 16 \mathcal{X}\mathcal{Z}
    + 17\mathcal{Z}^2\right) - \frac{2}{9} \mathcal{X} \left( \mathcal{X}^2 + 10\mathcal{X}\mathcal{Z} + \mathcal{Z}^2 \right),\\
  C_4(\mathcal{X},\mathcal{Z}) &= -\frac{56}{3} \mathcal{X}
  + \frac{2\sqrt{5}}{9}\left( \mathcal{X}^2 + 10\mathcal{X}\mathcal{Z} + \mathcal{Z}^2 \right), \\[2mm]
  C_5(\mathcal{X},\mathcal{Z}) &= 0.
\end{align*}

The field strength of $A$, $F_{mnpq},$ and $F^{mnpq}$ also share the antisymmetry property of $A_{mnp}$ under the interchange of tensors defined from $Y^{\pm}_{IJKL}$ and $Z^{\pm}.$

This allows us to derive an expression for
\begin{align}
  \Delta^{-1}F_{mpqr}&F^{npqr} \notag \\
&\hspace{-10mm} = -6 \Delta^{-1}R_m{}^n + \frac{4 \,m_7^2 \, \Delta^{6}}{81} \bigg( 14\mathcal{X}^4 + 35\mathcal{X}^3\mathcal{Z} + 178\mathcal{X}^2\mathcal{Z}^2 +
  35\mathcal{X}\mathcal{Z}^3 + 14\mathcal{Z}^4 \nonumber\\
  &+ 3\sqrt{5}(\mathcal{X} + \mathcal{Z}) (19\mathcal{X}^2 - 50\mathcal{X}\mathcal{Z} + 19\mathcal{Z}^2) + \frac{63}{4}
  (29\mathcal{X}^2 - 190\mathcal{X}\mathcal{Z} + 29\mathcal{Z}^2) \bigg) \, \delta_m^n,
  \label{FFinv}
\end{align}
where we have used the expressions for $F_{mnpq}$ and $F^{mnpq}$, equations \eqref{F} and \eqref{Finv}, respectively, as well as equation \eqref{riccimn} and the SO(7) identities in section~\ref{sec:so7iden}.

Finally, contracting the indices in the equation above and using the expression for $R_{m}{}^{n}$ in equation \eqref{riccimn} as well the SO(7) identities gives
\begin{align}
  \Delta^{-1} F_{mnpq} F^{mnpq} &= \frac{16 m_7^2 \, \Delta^{6}}{27} \big( 14\mathcal{X}^4
  + 35\mathcal{X}^3\mathcal{Z} + 178\mathcal{X}^2\mathcal{Z}^2 + 35\mathcal{X}\mathcal{Z}^3 + 14\mathcal{Z}^4 \nonumber\\[2mm]
  &\quad + 3\sqrt{5}(\mathcal{X}+\mathcal{Z})(19\mathcal{X}^2 - 50\mathcal{X}\mathcal{Z} + 19\mathcal{Z}^2) -
  189(3\mathcal{X}-\mathcal{Z})(3\mathcal{Z}-\mathcal{X}) \big).
  \label{F2}
\end{align}

\subsection{The Einstein and Maxwell equations}
\label{sec:einsteins-equations}
Using  equations \eqref{ricci4d}, \eqref{riccimn}, \eqref{FFinv}, \eqref{F2} and \eqref{det}, it is now straightforward to show that the Einstein equations \eqref{Einseqn1} and \eqref{Einseqn2} are satisfied for the values of $m_4$ and $\fr$ given in \eqref{m4fr}. Finally, 
using the equations for the derivatives of the SO(7) tensors
\eqref{diffident} to differentiate \eqref{Finv} as well as the $\rm SO(7)$ identities in section~\ref{sec:so7iden}, we find that the Maxwell equation (\ref{maxwell})
is also satisfied.

\sect{Solution in ambient coordinates}
\label{sec:amb}
                                                                                                                                                                                                                                                                                                                                                                                                                                                                                                                                                                                            
The solution presented in the previous sections is given in terms of quantities defined on the round seven-sphere.  In particular, the metric, $g_{mn}$, in \eqref{soln:met} is written as a deformation of the metric, $\go_{mn}$, on the round seven-sphere. Furthermore, the tensors \eqref{so7tensors} are defined in terms of the Killing spinors on $S^7$. While this is necessary for obtaining the solution via the uplift ans\"atze \eqref{ans:metric} and \eqref{ans:flux} for the metric and flux, it is perhaps not the most natural form in which to express the solution given its isometry. In this section, we present
the solution in a form in which the action of the SO(3)$\times$SO(3) is more manifest.

\subsection{Ambient coordinates}
To find the relation between the coordinates on the round seven-sphere and
coordinates that we will use in this section, we introduce coordinates
$x^A$ on $\mathbb{R}^8$, where $A= 1, \dots, 8$. Then the seven-sphere is
defined by
\begin{equation}
 m_7^2 \, x\cdot x = 1,
 \label{7sphere}
\end{equation}
where in this section we use the notation $ x \cdot x \equiv x^{A}
x^{A}.$ It is straightforward to see that the above relation is solved by
\begin{align}
 m_7 \, x^{m} = \frac{2 y^{m}}{1 + |y|^2}, \qquad m_7 \, 
 x^8 = \frac{1- |y|^2}{1+ |y|^2},
 \label{ambstercoor}
\end{align}
which define stereographic coordinates $y^m$ on the round seven-sphere of inverse
radius $m_7$ (with $|y|^2\equiv y^m y^m$). The relations in the previous section can be viewed
as being written in precisely such a coordinate system.  Hence, in the
previous sections the line element on the round $S^7$ is given by
\begin{equation}
 d s^{2} = \go_{mn} \,d y^{m} d y^{n}.
\end{equation}
In fact, the induced metric on the seven-sphere can easily be calculated by
substituting equations \eqref{ambstercoor} into the flat line element on
$\mathbb{R}^8$, whereupon we find that 
\begin{equation}
 \go_{mn} = \frac{4}{m_7^2 (1+ |y|^2)^2} \delta_{mn}.
\end{equation}
A convenient choice for the siebenbein is
\begin{equation}
 \eo^{a} = - \frac{2}{m_7 (1+ |y|^2)} d y^a.
\end{equation}

Instead of viewing the action of SO(3)$\times$SO(3) in stereographic
coordinates, we can now view its action as an action of SO(3)$\times$SO(3) $\simeq$ SO(4) 
on two four-dimensional subspaces of $\mathbb{R}^8$ in ambient coordinates $x^{A}$. 
More precisely, we can view $\mathbb{R}^8$ as the direct sum $\mathbb{R}^4
\oplus \mathbb{R}^4$ and decompose $x= (u,v)$, where $u,v
\in\mathbb{R}^4$ such that SO(4) acts separately on $u$ and $v$.~\footnote{No confusion should arise between these and the $u$ and $v$ matrices that parametrise the scalars in the gauged theory, described in section~\ref{sec:gaugedsoln}.}

The SO(3)$\times$SO(3) invariant tensors in the previous section,
written in terms of Killing spinors on the round $S^7$, can be expressed
in ambient coordinates as follows. In terms of Killing spinors, the
1-form duals of Killing vectors on $S^{7}$ are \cite{dWNW}
\begin{equation}
 K^{IJ} = K_{a}^{IJ} \eo^{a}.
\end{equation}
However, since the Killing vectors, $K_a^{IJ}$, generate SO(8) in \textbf{28}, they are related by
triality to generators of SO(8) in the vector representation. Or,
equivalently, in terms of their 1-form duals
\begin{equation}
 K^{IJ} = -\frac{m_7}{2} \Gamma^{IJ}_{AB} \cK^{AB}, \qquad \cK^{AB} =
-\frac{1}{8 m_7} \Gamma^{AB}_{IJ} K^{IJ},
 \label{kijrelns}
 \end{equation}
 where
 \begin{equation}
  \cK^{AB} = 2 x^{[A} d x^{B]}.
  \label{ckdef}
 \end{equation}
Furthermore,
\begin{equation}
 K_{(2)}^{IJ} \equiv \frac{1}{2} \, K_{ab}^{IJ} \eo^{a} \wedge \eo^{b} =
\frac{1}{2} \, \Gamma^{IJ}_{AB} \, d \cK^{AB}.
 \label{ck2def}
\end{equation}
Now we can use these relations to determine the SO(3)$\times$SO(3)
invariant tensors in ambient coordinates. We start with the scalar
invariant $\xi$ defined in \eqref{so7tensors}, and substitute for $K_{m}^{IJ}$ using relation
\eqref{kijrelns}
\begin{equation}
 \xi = \frac{m_7^2}{16} \,  Y^{+}_{IJKL} \Gamma^{IJKL}_{AB} x^{A} x^{B}
= 3 \,  m_7^2 \, \Gamma^{1234}_{AB}\,  x^{A} x^{B}.
 \label{xisc}
\end{equation}
Note that since the exterior derivative in $\cK^{AB},$ definition
\eqref{ckdef}, is with respect to stereographic coordinates, we also
use relations \eqref{ambstercoor} in deriving the above result.
Similarly,
\begin{equation}
 \zeta = 3 \, m_7^2\,  \Gamma^{1235}_{AB} \, x^{A} x^{B}.
 \label{zetasc}
\end{equation}
Naively, there are three scalar invariants that can be formed from $u$
and $v$. However, note that from equation \eqref{7sphere}
\begin{equation}
  u\cdot u + v \cdot v =1.
  \label{uvs7id}
\end{equation}
Therefore, we only have two scalar invariants
$$ u\cdot u - v \cdot v, \qquad u \cdot v$$
and without loss of generality we can pick an embedding of the
$\mathbb{R}^4$ in $\mathbb{R}^{8}$ where
\begin{equation}
 \xi = - 3 (u\cdot u - v \cdot v), \qquad \zeta = -6 u \cdot v .
\label{xizetaam}
 \end{equation}
For an explicit embedding where the above relations hold see appendix
\ref{app:expliciteg}. Note that any other embedding will correspond to a
rotation between $u$ and $v$, which in the present representation, see
appendix~\ref{app:expliciteg}, is given by $\Gamma^{45}_{AB},$ \emph{viz}.\
\begin{equation}
 \Gamma^{45}: u \mapsto v,\ v \mapsto -u.
\end{equation}
This freedom is represented by the parameter $\alpha$ in section~\ref{sec:11dsoln},
which is related to the rotation angle between $u$ and $v.$ In the
four-dimensional theory, this corresponds to a redundancy in the
description of the  SO(3)$\times$SO(3) invariant stationary point and
not an invariance. As was shown in section~\ref{sec:alpha}, this is reflected in
the fact that the uplift of all these points correspond to the same
solution up to coordinate transformations.

Given the expressions for $\xi$ and $ \zeta$ in ambient coordinates, it
is now straightforward to find the tensors $\xi_{a}$ and $\zeta_{a}$ in
ambient coordinates by differentiating expressions \eqref{xizetaam} and
using equations in \eqref{diffident}:
\begin{equation}
 m_7 \, \xi_{a} \eo^{a} = - 3 \, (u \cdot d u -  v \cdot d v), \qquad
m_7\, \zeta_{a} \eo^{a} = - 3 \, (v \cdot d u + u \cdot d v).
\end{equation}
The remaining invariant 1-form $F_{a}$, \eqref{Ftens}, is found using equations
\eqref{kijrelns}, \eqref{ckdef} and the third equation in \eqref{4g2gid},
\begin{equation}
m_7 F_{a} \eo^{a} = v \cdot d u  - u \cdot d v.
\end{equation}

We may again differentiate the tensors $\xi_{a}$ and $\zeta_{a}$ to
obtain expressions for the symmetric tensors $\xi_{ab}$ and
$\zeta_{ab},$ respectively, in ambient coordinates. However, we will
instead find these expressions by other means, which will be applicable
also to the derivation of the tensors $S_{abc}$ and $T_{abc}.$ 

Using
equation \eqref{kijrelns}, we rewrite
\begin{equation}
 \xi_{ab} \, \eo^{a} \eo^{b} = - \frac{m_7^2}{64} \, Y^+_{IJKL}
\Gamma^{IJ}_{AB} \, \Gamma_{CD}^{KL} \, \cK^{AB} \, \cK^{CD}.
\end{equation}
Note that the indices on $Y^{+}$ fully antisymmetrise the indices on the
$\Gamma$-matrices. Hence we can make use of the following identity
\cite{so(8)}:
\begin{equation}
 \Gamma^{[IJ}_{AB} \, \Gamma^{KL]}_{CD} = \frac{1}{2} \left(
\Gamma^{IJ}_{[AB} \,  \Gamma^{KL}_{CD]} - \frac{1}{24}
\epsilon_{ABCDEFGH}  \, \Gamma^{IJ}_{EF} \, \Gamma^{KL}_{GH} \right) +
\frac{2}{3} \delta^{\phantom{I}}_{[C|[B} \, \Gamma^{IJKL}_{A]|D]},
 \label{triid}
\end{equation}
which is a consequence of SO(8) triality and is a decomposition of the
object on the left hand side into its anti-selfdual (first term) and
selfdual part (second term). Moreover, noting that in the expression for
$\xi_{ab}$ the combination of $\Gamma$-matrices contracts with a
selfdual tensor, $Y^{+}_{IJKL}$, we obtain
\begin{equation}
  \xi_{ab} \eo^{a} \eo^{b} = \frac{m_7^2}{2} \, \Gamma^{1234}_{AB} \,
\cK^{AC} \, \cK^{CB}.
\end{equation}
Finally using \eqref{ckdef} and the first equation in
\eqref{4g2gid}, we find that
\begin{equation}
  m_7^2 \, \xi_{ab} \, \eo^{a} \eo^{b} = (v \cdot v) \, d v \cdot d v
- (u \cdot u) \, d u \cdot d u,
\end{equation}
where we have also used
\begin{equation}
 u \cdot d u + v \cdot d v = 0,
\end{equation}
which follows from  \eqref{uvs7id}.
Similarly, we also find
\begin{equation}
 m_7^2 \, \zeta_{ab} \, \eo^{a} \eo^{b} = - \Big[ u \cdot v \, (d u
\cdot d u +d v \cdot d v) + d u \cdot d v \Big].
\end{equation}

We determine $S_{abc}$ and $T_{abc}$ in an
analogous way. For example,
\begin{equation}
  S_{(3)} \equiv \frac{1}{6} \, S_{abc} \,  \eo^{a} \eo^{b} \eo^{c} = -
  \frac{m_7}{192}\,  Y^{-}_{IJKL} \,  \Gamma^{IJ}_{AB}\, \Gamma^{KL}_{CD}
  \, \cK^{AB_{(2)}} \wedge \cK^{CD}.
 \end{equation}
Hence, we can again use identity \eqref{triid}, but in this case the
anti-selfdual part of the decomposition given in equation \eqref{triid}
survives and we obtain
\begin{equation}
 S_{(3)} = - \frac{m_7}{4} \, \Gamma^{12}_{AB} \, \Gamma^{35}_{CD} \,
x^{[A} d x^{B} \wedge d x^{C} \wedge d x^{D]},
\end{equation}
which can be evaluated using the $\Gamma$-matrices and the embedding
given in appendix~\ref{app:expliciteg}. All in all, we obtain
\begin{align}
 m_7^3 \, S_{(3)} &= - \frac{1}{12} \Big[ \epsilon(u, d v, d v, d v)
+ \epsilon(v, d u, d u, d u) + 3 \epsilon(u, d u, d u, d v) + 3
\epsilon(v, d u, d v, d v) \Big], \\[2mm]
  m_7^3 \, T_{(3)} &= - \frac{1}{6} \Big[ \epsilon(u, d u, d u, d u)
- \epsilon(v, d v, d v, d v) \Big],
\end{align}
where we have introduced the convenient notation
\begin{equation}
 \epsilon(u, d u, d u, d v) \equiv \epsilon_{ijkl} \, u^{i} \, d u^{j} \wedge d u^{k} \wedge d v^{l}.
\end{equation}
It is clear that there are two more invariant 3-forms,
\begin{equation}
 \epsilon(u, d u, d v, d v), \qquad \epsilon(v, d u, d u, d v),
\end{equation}
that do not appear in the expression for $S_{(3)}$ or
$T_{(3)}.$ However, these invariant 3-forms as well as the
3-forms in $S_{(3)}$ and $ T_{(3)}$ do appear in
the expression for the internal 3-form potential given below. 

\subsection{The solution}

In terms of the ambient coordinates introduced above, the solution \eqref{soln:met}-\eqref{soln:3form} reads:\footnote{K.P.\ would like to thank N.\ Bobev, A.\ Kundu and N.\ Warner for a collaboration which independently led to the metric in the ambient form presented here \cite{BKPW}.}
\begin{equation}
\begin{split}
ds_7^2  = \frac{\Delta^2}{6\,m_7^2}\Big[
\,& +c (6 c-s (\zeta +\xi ))\,(du\cdot du+dv\cdot dv)
 +s (s\xi  -3 c)\,(du\cdot du-dv\cdot dv)\\ 
& +2 s (s\zeta  -3 c)\,du\cdot dv+\frac{1}{6}\,s^2 \,f^2\Big]\,,
 \label{soln:metam}
\end{split}
\end{equation}
and
\begin{equation}\label{soln:3formam}
\begin{split}
A_{(3)}  = \frac{\sqrt 2}{144}  \frac{\Delta^3}{m_7^3}\,
 \Big[\,
  & +s \left(12 c^3-c^2 s (2 \zeta +2 \xi +3)-c s^2 (\zeta -2 \xi +3)+\zeta s^3\right)
   \epsilon(u, {d u}, {d u}, {d u}) \\[-3 pt]
   & -s \left(12 c^3+c^2 s (-2 \zeta -2 \xi +3)+c s^2 (\zeta -2\xi -3)+\zeta  s^3\right)
   \epsilon (v, {d v}, {d v}, {d v}) \\[3 pt]
& -3 s (c+s) \left(6 c^2-c s (\zeta +\xi +3)+\xi  s^2\right) \epsilon (u, {d u}, {d u}, {d v})\\[3 pt]
&  -3 s (c-s)\left(6 c^2-c s (\zeta +\xi -3)-\xi  s^2\right) \epsilon (v, {d u}, {d v}, {d v}) \\[3 pt]
   & -s (c-s) \left(6 c^2-c s (\zeta +\xi +3)+\xi  s^2\right)\epsilon(v, {d u}, {d u}, {d u}) \\[3 pt]
   & +s (c+s) \left(-6 c^2+c s (\zeta +\xi -3)+\xi  s^2\right)\epsilon(u, {d v}, {d v}, {d v}) \\[3 pt] 
    &-3 s^2 (c+s) (\zeta  s-3 c) \epsilon (u, {d u}, {d v}, {d v})\\[3 pt]
   &+3 s^2 (c-s) (3c-\zeta  s) \epsilon (v, {d u}, {d u}, {d v})\,\Big]\,,
\end{split}
\end{equation}
where
\begin{equation}\label{}
ds_7^2\eql g_{ab}\,\eo^a\eo^b\,,\qquad A_{(3)}\eql \frac{1}{6}\,A_{abc}\eo^a\wedge \eo^b\wedge \eo^c\,,
\end{equation}
\begin{equation}\label{vvect}
m_7\,f\equiv m_7\,f_a \eo^a =  3c(u\cdot du-v\cdot du-u\cdot dv-v\cdot dv)+3s(v\cdot du-u\cdot dv)\,
\end{equation}
and with $c$ and $s$ set to their stationary values \eqref{statval}.

\subsection{Local coordinates}
We conclude this section with a construction of local coordinates on $S^7$ using the Euler angles of the $\rm SO(3)\times SO(3)$ isometry group and the two scalar invariants, $\xi$ and $\zeta$. To this end let us consider $S^7$ as   a subspace of $2\times 2$ complex matrices
\begin{equation}\label{Zmatrix}
{\mathbf{Z}}\eql \left(\begin{matrix}
z^3-i z^2 & z^1+i z^4 \\ -z^1+ i z^4 & z^3+iz^2
\end{matrix}\right)\,,\qquad z^j\eql u^j+iv^j\,,\end{equation}
satisfying
\begin{equation}\label{eq:tracecond}
\frac{1}{2}\,
{\rm Tr}\, {\mathbf{Z}}{\mathbf{Z}}^\dagger =u\cdot u+v\cdot v\eql  1\,
\end{equation}
and
\begin{equation}\label{eq:detcond}
\det\,{\mathbf{Z}}= -\frac{1}{3}\,(\xi+i\zeta)\,.
\end{equation}
Then the $\rm SO(4)$ action on $\mathbb C^4$ is the same as the action of $\rm SU(2)_1\times SU(2)_2$ on such matrices given by 
\begin{equation}\label{}
{\cal Z}\quad\longrightarrow\quad  \cR_1\,{\cal Z}\,\cR_2^\dagger\,,
\end{equation}
under which both \eqref{eq:tracecond} and \eqref{eq:detcond} remain invariant.

We use the Euler angles for the two $\rm SU(2)$s defined by
\begin{equation}\label{}
\cR_j(\theta_j,\phi_j,\psi_j)\eql \left(\begin{matrix}
e^{\frac{i}{2}(\phi_j+\psi_j)}\cos{\frac{\theta_j}{2}} & -e^{\frac{i}{2}(\phi_j-\psi_j)}\sin{\frac{\theta_j}{2}}\\ e^{-\frac{i}{2}(\phi_j-\psi_j)}\sin{\frac{\theta_j}{2}} & e^{-\frac{i}{2}(\phi_j+\psi_j)}\cos{\frac{\theta_j}{2}}
\end{matrix}\right)\,,\qquad j=1,2\,.
\end{equation}
By an $\rm SU(2)_1\times SU(2)_2$ transformation, one can bring $\mathbf Z$ to a diagonal form,
\begin{equation}\label{diagpar}
{\mathbf{Z}}_d(\rho,\varphi)\eql \sqrt 2\,e^{\frac{i}{2} {(\varphi+\pi)}}\,\left(\begin{matrix}
\cos{\frac{\rho}{2}} & 0 \\ 0 & \sin{\frac{\rho}{2}}
\end{matrix}\right)\,,\qquad 0\leq \rho\leq {\frac{\pi}{2}}\,,\quad 0\leq \varphi\leq 2\pi\,,
\end{equation}
where $(\rho,\varphi)$ parametrise a disk of radius $\pi/2$. Using \eqref{eq:detcond}, we find 
\begin{equation}\label{xzrho}
\xi\eql 3\,\sin\rho\cos\varphi\,,\qquad \zeta\eql 3\sin\rho\sin\varphi
\end{equation}
so that have  $|\xi|,|\zeta|\leq 3$, which is consistent with identities (i) in table~\ref{so7id1} \cite{dWNso7soln}.

At a generic point,  we  have
\begin{equation}\label{decomp}
{\mathbf{Z}}\eql \cR_1(\theta_1,\phi_1,\psi_1)\,{\cal Z}_d(\rho,\varphi)\,\cR_2(\theta_2,\phi_2,\psi_2)^\dagger\,.
\end{equation}
Clearly,  $\mathbf{Z}$ is invariant under $\psi_i\rightarrow\psi_i+\chi$, which shows that a typical orbit is isomorphic with the coset 
\begin{equation}\label{}
 \frac{SU(2)_1\times SU(2)_2}{U(1)}\,,
\end{equation}
where $U(1)$ is the diagonal subgroup.\footnote{Note that at the center of the disk $\xi=\zeta=0$ and we simply reproduce the explicit construction of $T^{1,1}$ in \cite{Candelas:1989js}.} The local coordinate system on $S^7$ is now comprised of the angles $\rho$ and $\phi$ that parametrise a disk and the Euler angles  $\theta_1,\phi_1,\theta_2,\phi_2$ and $\psi=\psi_1-\psi_2$ on the coset. The range of these angles are
\begin{equation}\label{}
\begin{split}
0\leq \rho\leq \frac{\pi}{2}\,,\qquad  0\leq \theta_2\,,\psi\leq \pi\,,\qquad
 0\leq \varphi\,, \phi_1,\phi_2,\theta_1\leq 2\pi\,.
\end{split}
\end{equation}

Let us also introduce the left invariant forms on $SU(2)_1\times SU(2)_2$,
\begin{equation}\label{thesigmas}
\begin{split}
\sigma_{(j)}^1 & \eql \sin\psi_j\,d\theta_j-\cos\psi_j\sin\theta_j\,d\phi_j\,,\\
\sigma^2_{(j)} & \eql -\cos\psi_j\,d\theta_j-\sin\psi_j\sin\theta_j\,d\phi_j\,,\\
\sigma_{(j)}^3 & \eql d\psi_j+\cos\theta_j\,d\phi_j\,,\\
\end{split}
\end{equation}
satisfying $d\sigma_{(j)}^1=\sigma_{(j)}^2\wedge\sigma_{(j)}^3$, etc.,
and define
\begin{equation}\label{}
\sigma_{(j)}^\pm\eql \sigma_{(j)}^1\pm i\sigma^2_{(j)}\,.
\end{equation}
These forms are then pulled-back onto the coset by setting $\psi_1=-\psi_2\eql\psi/2$, such that   
\begin{equation}\label{}
\sigma^1_{(1)}\,,\quad \sigma^2_{(1)}\,,\quad \sigma^1_{(2)}\,,\quad \sigma^2_{(2)}\,,\quad \sigma^3\equiv\sigma^3_{(1)}-\sigma^3_{(2)}\,,
\end{equation}
yield a local frame, $\sigma^a$, $a=1,\ldots,5$, along the orbits of the $\rm SO(4)$ isometry. 

The round metric on $S^7$ in these coordinates reads
\begin{equation}\label{}
d\mathring s ^2_7 = {\frac{1}{4m_7^2}}\,\Big[ d\rho^2+\sin^2\rho\,d\varphi^2 +\big(\sigma_{(1)}^+\sigma_{(1)}^-+\sigma_{(2)}^+\sigma_{(2)}^-\big)
-\sin\rho\,\big(\sigma_{(1)}^+\sigma_{(2)}^-+\sigma_{(2)}^+\sigma_{(1)}^-\big)+\big(\cos\rho\,d\varphi-\sigma^3\big)^2\,
\Big]\,.
\end{equation}
The  geometric objects  \eqref{ourtens} are the scalars given by \eqref{xzrho}, the vectors:  
\begin{equation}\label{}
\begin{split}
m_7(\cos\varphi \,\xi_a+\sin\varphi\, \zeta_a)\eo^a &  \eql {\frac{3}{2}}\cos\rho\,d\rho\,\,,\\[6 pt]
m_7(\cos\varphi \,\zeta_a -\sin\varphi\, \xi_a)\eo^a  & \eql {\frac{3}{2}}\sin\rho\,d\varphi\,,
\end{split}
\end{equation}
the  symmetric tensors:
\begin{equation}\label{}
\begin{split}
m_7(\cos\varphi \,\xi_{ab}+\sin\varphi\, \zeta_{ab})\eo^a\eo^b &\eql -{\frac{1}{4}} \sin\rho\big(\cos\rho\,d\varphi -\sigma^3\big)\sigma^3 +
{\frac{1}{4}}\,\sin\rho\,\big(\sigma_{(1)}^+\sigma_{(1)}^-+\sigma_{(2)}^+\sigma_{(2)}^-\big)\\[6 pt]
&\qquad\quad  +{\frac{1}{16}}\,  
(\cos(2\rho)-3)\big(\sigma_{(1)}^+\sigma_{(2)}^-+\sigma_{(2)}^+\sigma_{(1)}^-\big)\,,
\\[6 pt]
m_7(\cos\varphi \,\zeta_{ab} -\sin\varphi\, \xi_{ab})\eo^a\eo^b  &\eql {\frac{1}{4}}\,d\rho\,(\cos\rho\,d\varphi-\sigma^3)+{\frac{i}{8}}\,\cos\rho\big( \sigma_{(1)}^+\sigma_{(2)}^--\sigma_{(2)}^+\sigma_{(1)}^-\big)
\,,
\end{split}
\end{equation}
and the 3-forms:
\begin{equation}\label{}
\begin{split}
{m_7^3}\,& \big(\cos\varphi\, S_{abc}- \sin\varphi \,T_{abc}\big)\,\eo^a\wedge\eo^b\wedge \eo^c\eql {\frac{3}{16}}\times \\[6 pt] &
\Big\{-i\,d\rho\wedge\Big[\big(\sigma^+_{(1)}\wedge\sigma^-_{(1)}+\sigma^+_{(2)}\wedge\sigma^-_{(1)}\big)-\sin\rho\,
\big(\sigma^+_{(1)}\wedge\sigma^-_{(2)}+\sigma^+_{(2)}\wedge\sigma^-_{(1)}\big)\Big]\\[6 pt]
& \hspace{5.2cm}+\cos\rho\,( \cos\rho\,d\varphi-\sigma^3)\wedge  \big(\sigma^+_{(1)}\wedge\sigma^-_{(2)}-\sigma^+_{(2)}\wedge\sigma^-_{(1)}\big)\Big\}\\[6 pt]
{m_7^3}\,&\big(\cos\varphi\, T_{abc}+ \sin\varphi \,S_{abc}\big)\,\eo^a\wedge\eo^b\wedge \eo^c\eql {\frac{3i}{16}}\,\times\\[6 pt] &
\Big\{(\cos\rho\,d\varphi-\sigma^3)\wedge \Big[
\big(\sigma^+_{(1)}\wedge\sigma^-_{(2)}+\sigma^+_{(2)}\wedge\sigma^-_{(1)}\big)
-\sin\rho\,
\big(\sigma^+_{(1)}\wedge\sigma^-_{(1)}+\sigma^+_{(2)}\wedge\sigma^-_{(1)}\big)\Big]\\[6 pt]
&\qquad\qquad   +\sin\rho\,\sigma^3\wedge\Big[ 
\big(\sigma^+_{(1)}\wedge\sigma^-_{(1)}+\sigma^+_{(2)}\wedge\sigma^-_{(1)}\big)-\sin\rho\,
\big(\sigma^+_{(1)}\wedge\sigma^-_{(2)}+\sigma^+_{(2)}\wedge\sigma^-_{(1)}\big)
\Big]\Big\}\,.
\end{split}
\end{equation}
We also have that
\begin{equation}
\begin{split}
m_7\,F_a \eo^a  & = -{\frac{1}{2}} \big(d\varphi -\cos\rho\, \sigma^3)\,,\\
m_7^2\,F_{ab} \eo^a\wedge \eo^b  & = -\frac{1}{2}\sin\rho\, d\rho\wedge\sigma^3- {\frac{i}{4}}\cos\rho\,\big(\sigma^+_{(1)}\wedge \sigma^-_{(1)}-\sigma^+_{(2)}\wedge \sigma^-_{(2)} \big) \,.\\
\end{split}
\end{equation}
Rotations by the angle $\varphi$ to obtain the actual $\rm SO(7)$ tensors  \eqref{ourtens}  result in even larger expressions. As expected, the explicit formulae for the metric \eqref{soln:met} and the 3-form potential  \eqref{soln:3form} in these local coordinates  are quite  complicated and we will not write them here. One can easily obtain them using the expressions for the $\rm SO(7)$ tensors given above.

\sect{Outlook}
\label{sec:conclusions}

In this paper, we have constructed a new and highly non-trivial solution of $D=11$ supergravity corresponding to an uplifting of the SO(3)$\times$SO(3) invariant stationary point of maximal gauged supergravity.  While this solution is of interest in holographic applications and we hope that readers will find good use for it, we have endeavored to present the derivation of the solution in such a manner as to lend itself to a more general explanation of uplifting solutions of this type, i.e. Freund-Rubin compactifications with internal flux.  The uplifting of any stationary point of the gauged theory to eleven dimensions will follow the same steps as those presented for the SO(3)$\times$SO(3) invariant stationary point here, except that, clearly, for stationary points with less symmetry, this will be a more cumbersome process with many different invariant forms to consider.

Apart from allowing for a direct derivation of uplift formulae, the rewriting of the eleven-dimensional theory in an SU(8) invariant reformulation \cite{dWNsu8}, highlights features of the four-dimensional theory in eleven dimensions and makes it possible to prove \cite{dWNconsist87, NP}, for example, the consistency of the $S^7$ reduction \cite{freundrubin, duffpope}.

In recent work \cite{GGN13,GGN14}, the ideas initiated in Ref.~\cite{dWNsu8} are taken to their full conclusion giving an on-shell equivalent reformulation of the $D=11$ theory in which features of the global group E$_{7(7)}$ are also made manifest.  As well as breaking manifest eleven-dimensional Lorentz invariance and covariance, one is also compelled to introduce eleven-dimensional dual fields in order to bring out the E$_{7(7)}$ structure.  

The reformulation of $D=11$ supergravity given in Ref.~\cite{GGN13} provides a very direct and efficient way of studying the relation between four-dimensional maximal gauged theories and $D=11$ supergravity via a higher-dimensional understanding \cite{GGN13} of the embedding tensor \cite{NSmaximal3, NScomgauge3, dWSTlag, dWSTmax4}.  In particular, it allows for a simple analysis of which four-dimensional theories arise as consistent reductions of the eleven-dimensional theory (see e.g.~\cite{GGNss}). For example, it is very simple to deduce \cite{KKdual} that the new deformed SO(8) gauged theories of Ref.~\cite{DIT, DIM} cannot be obtained from a consistent reduction of the $D=11$ theory. 

In fact, given the success of the reformulations described above, we argue that, generally, the most appropriate setting in which to address questions to do with reductions and consistency is one in which the higher-dimensional theory is reformulated in such a manner as to fully resemble a duality covariant reformulation of the lower-dimensional theory, including both the global and local duality groups. 

Of particular relevance here is that in the case of the $S^7$ reduction to the original maximal SO(8) gauged theory \cite{dWNn8}, Ref.~\cite{GGN13} completes the metric and flux ans\"atze and provides full uplift ans\"atze for \emph{any} solution of the gauged theory to eleven dimensions, including dynamical solutions with non-trivial $x$-dependence \cite{KKdual}. The method can, however, be applied more generally. For example, one can in principle setup a reformulation along the lines of \cite{dWNsu8, GGN13} for type IIB supergravity and thereby study its $S^5$ truncation---for a recent conjecture on uplift ans\"atze in this case see Ref.~\cite{Lee:2014mla}.  

An interesting application of these full uplift ans\"atze \cite{GGN13, KKdual} would be to construct the full interpolating solution for a particular RG flow between two stationary points of the potential, such as the flow between the maximally symmetric SO(8) and the SO(3)$\times$SO(3) invariant stationary points considered in Ref.~\cite{Fischbacher:2010ec}.

\vspace{1cm}

\noindent\textbf{Acknowledgements}:
We are grateful to Nikolay Bobev, Arnab Kundu, Chris Pope, Harvey Reall and Nick Warner for discussions. M.G., H.G.~and K.P.~ would like to thank the Max-Planck-Institut f\"ur Gravitationsphysik (AEI) and in particular H.N. for hospitality. H.G.\ and M.G.\ are supported by King's College, Cambridge. H.G.\ acknowledges funding from the European Research Council under the European Community's Seventh Framework Programme (FP7/2007-2013) / ERC grant agreement no. [247252]. K.P.~was supported in part by DOE grant DE-SC0011687.

\newpage

\appendix
\section{Conventions}
\label{app:conventions}

\renewcommand{\theequation}{A.\arabic{equation}}
\setcounter{equation}{0}

We define a set of euclidean, antisymmetric and purely imaginary
$8\times8$ $\Gamma$-matrices ($\Gamma^\dagger = \Gamma$). These are
generators of the euclidean Clifford algebra in seven dimensions,
\begin{eqnarray}
  \label{eq:149}
  \{\Gamma_a,\Gamma_b \} = 2 \delta_{ab} \mathbb{I}_{8\times8}.
\end{eqnarray}
We choose a Majorana representation and set the charge conjugation
matrix that defines spinor conjugates or raises and lowers spinor
indices to be the unit matrix. An explicit representation for the $\Gamma$-matrices is given in appendix~\ref{app:expliciteg}.

The $\Gamma$-matrices can be used to define the $8\times8$ matrices
\begin{eqnarray}
  \label{eq:150}
  \Gamma_{a_1 \ldots a_i} = \Gamma_{[ a_1} \ldots \Gamma_{a_i]}
\end{eqnarray}
for $i = 2, \ldots 7$. $\Gamma_a$ and $\Gamma_{ab}$ are antisymmetric
matrices and $\Gamma_{abc}$ is symmetric. These $7 + 21 + 35 = 63$
matrices together with the unit matrix span the vector space of
$8\times8$ matrices. Thus, we find that
\begin{eqnarray}
  \label{eq:151}
  \Gamma_{a_1 \ldots a_7} &=& -i \eta_{a_1 \ldots a_7},\\
  \Gamma_{a_1 \ldots a_6} &=& -i \eta_{a_1 \ldots a_6 b}
  \Gamma_{b},\\ 
  \Gamma_{a_1 \ldots a_5} &=& \frac{i}{2} \eta_{a_1 \ldots a_5
    bc} \Gamma_{bc},\\
  \Gamma_{a_1 \ldots a_4} &=& \frac{i}{3!} \eta_{a_1 \ldots a_4
    bcd} \Gamma_{bcd}.
\end{eqnarray}
Furthermore, it is useful to note that each product of
$\Gamma$-matrices can be written in terms of the unit matrix,
$\Gamma_a$, $\Gamma_{ab}$ and $\Gamma_{abc}$.

We choose the eight Killing spinors of the round $S^7$ to be orthonormal,
\begin{eqnarray}
  \label{orthonormality}
  \bar \eta^I \eta^J = \delta^{IJ}, \quad \eta^I \bar \eta^I =
  \mathbb{I}_{8\times8},
\end{eqnarray}
where $\bar \eta^I = (\eta^I)^\dagger$.

The curved $\Gamma$-matrices on the
round seven-sphere are given by $\mathring \Gamma_m = \mathring e_m{}^a \Gamma_a$.
Hence, in our conventions, the Killing spinors satisfy
\begin{eqnarray}
  \label{killspineqn}
  i \mathring D_m \eta^I = \frac{m_7}{2} \mathring \Gamma_m
  \eta^I.
\end{eqnarray}

The Killing spinors define a set of Killing vectors, 2-forms and tensors:
\begin{eqnarray}
  \label{killspingamma}
  K_m^{IJ} = i \bar \eta^I \mathring \Gamma_m \eta^J, \quad
  K_{mn}^{IJ} = \bar \eta^I \mathring \Gamma_{mn} \eta^J, \quad K_{mnp}^{IJ} = i \bar \eta^I \mathring \Gamma_{mnp} \eta^J,
\end{eqnarray}
respectively, whose equivalents are also defined in flat space. Using
equation \eqref{killspineqn}, the reader may check that $K_{mn}^{IJ}$ is
proportional to the derivative of $K_m^{IJ}$,
\begin{eqnarray}
  \label{eq:155}
  \mathring D_n K_m^{IJ} = m_7 K_{mn}^{IJ}, \quad \mathring D_p K_{mn}^{IJ}
  = 2 m_7 \mathring g_{p[m} K_{n]}^{IJ}.
\end{eqnarray}
Note that curved seven-dimensional indices of the Killing vectors and
their derivatives are raised and lowered with the round seven-sphere
metric $\mathring g_{mn}$. 

As all $\Gamma$-matrices are traceless, we find that
\begin{eqnarray}
  \label{trace}
  \bar \eta^I \mathring \Gamma_{m_1\ldots m_i} \eta^I = 0 
\end{eqnarray}
for $i = 1,\ldots,6$.

\section{Derivation of SO(7) tensor identities}
 \label{app:idproofs}

\renewcommand{\theequation}{B.\arabic{equation}}
\setcounter{equation}{0}

In this appendix, we sketch the derivation of the $\rm SO(7)$ identities, listed in tables~\hbox{\ref{so7id1}-\ref{so7id7}}, for the SO(3)$\times$SO(3)
invariant tensors \eqref{ourtens} .

In the derivations below, we make heavy use of the completeness relation 
 \begin{equation} \label{complete}
   16\, \delta_{IJ}^{KL} = 2 K_m^{IJ} K^{m\, KL} + K_{mn}^{IJ} K^{mn \, KL}\,,
 \end{equation}
as well as  the following useful identities \cite{dWNsu8}:
\begin{gather}\label{yzcontract1}
 \frac{1}{16} Y^+_{IJKL} K_{mn}^{IJ} K_{p}^{KL} = -\frac{1}{3} \go_{p[m} \xi_{n]}, \qquad
 \frac{1}{16} Y^+_{IJKL} K_{mn}^{IJ} K^{pq\, KL} = -4\,
 \delta^{[p}_{[m} \xi^{\phantom{[p}}_{n]}{}^{q]} + \frac{2}{3} \xi \delta_{mn}^{pq},
 \\[2mm]
 \label{yzcontract2}
  \frac{1}{16} Z^+_{IJKL} K_{mn}^{IJ} K_{p}^{KL} = -\frac{1}{3} \go_{p[m} \zeta_{n]}, \qquad
 \frac{1}{16} Z^+_{IJKL} K_{mn}^{IJ} K^{pq\, KL} = -4\,
 \delta^{[p}_{[m} \zeta^{\phantom{[p}}_{n]}{}^{q]} + \frac{2}{3} \zeta
 \delta_{mn}^{pq}, \\[2mm]
 \label{yzcontract3}
  \frac{1}{16} Y^-_{IJKL} K_{mn}^{IJ} K_{pq}^{KL} = -\frac{1}{6} \etao_{mnpqrst} S^{rst}, \qquad
    \frac{1}{16} Z^-_{IJKL} K_{mn}^{IJ} K_{pq}^{KL} = -\frac{1}{6} \etao_{mnpqrst} T^{rst}.
\end{gather}
\label{st2}
One can verify these using the inversion formulae \eqref{invertedid}.

\addtocontents{toc}{\protect\setcounter{tocdepth}{1}}

\subsection{Derivation of the identities in table~\ref{so7id1}}

\noindent \emph{\underline{Identities (i) and (ii)}}$\quad$ Consider the first equation in \eqref{yzeqn1} contracted with $K_t^{IJ}K^{t\,KL}$:
\begin{equation}
 K_t^{IJ} Y^+_{IJMN} (2 K_m^{MN} K^{m\, PQ} + K_{mn}^{MN} K^{mn \, PQ}) Y^+_{PQKL} K^{t\,KL} = K_t^{IJ} Z^-_{IJMN} K_{mn}^{MN} K^{mn \, PQ} Z^-_{PQKL} K^{t\,KL},
\end{equation}
where we have used the completeness relation \eqref{complete} and the fact that $K_m^{IJ}K_n^{KL} Z^-_{IJKL}=0$ by virtue of the fact that $K_m^{[IJ}K_n^{KL]}$ is selfdual, while $Z^-_{IJKL}$ is anti-selfdual.  Now, substituting for the SO(7) tensors using the definitions \eqref{so7tensors} and equation \eqref{yzcontract1} gives
\begin{equation} 
2 \xi^{mn}\xi_{mn} +\frac{1}{9} \go_{t[m}\xi_{n]} \go^{t[m}\xi^{n]} = T^{mnp} T_{mnp},
\end{equation}
which simplifies to
\begin{equation} \label{xixi1}
 \xi^m \xi_m + 6\xi^{mn}\xi_{mn} = 3 T^{mnp} T_{mnp}.
\end{equation}
Repeating the above steps, except now contracting the first equation in \eqref{yzeqn1} with $K_{tu}^{IJ}K^{tu\,KL}$ gives
\begin{equation}\label{xixi2}
  -4 \xi^2 + \xi^m \xi_m + 30\xi^{mn}\xi_{mn} = 9 T^{mnp} T_{mnp}
\end{equation}
Finally, by contracting the first cubic identity \eqref{eqtr1} with $K_{tu}^{IJ}K^{u\,KL}$ and simplifying as before, except that the completeness relation \eqref{complete} must be used twice, gives
\begin{equation}
  \left(36 + 2 \xi^2 - \xi^m \xi_m - 18 \xi^{mn} \xi_{mn} \right) \xi_t = 0.
\end{equation}
There are seemingly two cases to consider: first we consider the case in which the expression in the brackets vanishes.  Together, with equations \eqref{xixi1} and \eqref{xixi2}, we obtain the equations for $\xi^m\xi_m$, $\xi^{mn}\xi_{mn}$ and $T^{mnp}T_{mnp}$ in terms of $\xi^2$, as they appear in equations in (i) and (ii) in table~\ref{so7id1}. The equations derived from considering the second case, $\xi_m \equiv 0$, are already contained in equations (i) and (ii). However, in our case, $\xi_m  \not\equiv  0$ anyway.

Note that we had to use a cubic identity, \eqref{eqtr1}, to derive a quadratic identity. This seems strange and one may wonder whether that was necessary or whether the identity could have been derived from quadratic identities. However, a simple counting of the number of quadratic identities available gives two, whereas the number of unknown quantities that we have expressed in terms of $\xi^2$ is three. Note, however, that \eqref{eqtr1} is not used anymore in deriving the identities in table~\ref{so7id1}.

Interchanging $Y$ and $Z$ in the discussion above, or equivalently by considering the second identities in \eqref{yzeqn1} and \eqref{eqtr1} gives analogous expressions for $\zeta^m\zeta_m$, $\zeta^{mn}\zeta_{mn}$ and $S^{mnp}S_{mnp}$. 

\smallskip

\noindent \emph{\underline{Identities (iii) and (vi)}}$\quad$
This case is similar to the example above. We contract equations \eqref{yzeqn1} with $K^{IJ}_{mn} K^{p KL}$.
This gives identity (vi). Identity (iii) is obtained upon letting index $p=n$ and noting that the wedge product of an odd-form with itself vanishes, e.g.\
\begin{equation}
 \etao_{mnpqrst} S^{npq} S^{rst} =0.
\end{equation}
\noindent \emph{\underline{Identities (iv) and (v)}}$\quad$
These identities are derived by contracting equations \eqref{yzeqn1} with $K^{IJ}_{m} K_{n}^{KL}$ and $K^{IJ}_{m p} K_{n}{}^{p \,KL}$. Identities (i)--(iii) are used to simplify the expressions.
\smallskip

\noindent \emph{\underline{Identities (vii) and (viii)}}$\quad$
Contract identities \eqref{yzeqn1} with $K^{IJ}_{mn} K_{pq}^{KL}$ and use identities (i)--(v) to simplify.

\subsection{Derivation of the identities in table~\ref{so7id2}}
\noindent \emph{\underline{Identity (i)}}$\quad$
The third identity in the line is proved by contracting the last equality in \eqref{yzeqn2} by $\delta_{IJ}$. Using the appropriate inversion formulae in \eqref{invertedid} and 
\begin{equation}
  K_{[ab}^{[IJ} K_{c]}^{KL]} K_{de}^{[IJ} K_{f}^{KL]} = 32 \delta^{abc}_{def},
\end{equation}
we immediately find $S_{mnp} T^{mnp} =0.$

The first two identities are derived by contracting either equation in \eqref{yzeqn5} with $K_{m}^{IJ}K^{m\,KL}$ and $K_{mn}^{IJ}K^{mn\,KL}$.
\smallskip

\noindent \emph{\underline{Identity (ii)}}$\quad$
Contract the last equality in \eqref{yzeqn2} with $K^{m \,IJ}$, whereupon we find
\begin{align} \label{t2iide1}
 Z^{-}_{IKLM} Y^{-}_{JKLM} K^{m\,IJ} = 12 F^{m}.
\end{align}
We then make use of the inversion formula for  $Z^{-}_{IKLM}$, \eqref{invertedid} to find 
\begin{equation}
 Z^{-}_{IKLM} K^{m\,IJ} = \frac{1}{4} S^{npq} K_{np}^{[LM} K_{q}{}^{m \, J|K]}.
\end{equation}
Substituting this expression and the inversion formula for $Y^{-}_{JKLM}$ in equation \eqref{t2iide1}, gives the required result. 
\smallskip

\noindent \emph{\underline{Identity (iii)}}$\quad$
These are obtained by contracting identity \eqref{yzeqn5} with $K^{IJ}_{mn} K^{n KL}.$
\medskip

\noindent \emph{\underline{Identities (iv) and (v)}}$\quad$
The symmetric, in indices $m$ and $n$, part of these are derived by contracting \eqref{yzeqn5} with $K^{m\,IJ} K^{n \, KL}$ and  $K^{mp \,IJ} K^{n}{}_{p}{}^{KL}.$ The antisymmetric part is derived by contracting \eqref{yzeqn2} with $K^{mn \,IJ}$,
\begin{equation}
 \label{t2ivvde1}
  Z^{+}_{IKLM} Y^{+}_{JKLM} K^{mn\,IJ}= 12 F^{mn}, \qquad  Z^{-}_{IKLM} Y^{-}_{JKLM} K^{mn\,IJ} = 12 F^{mn}.
\end{equation}
The evaluation of the left hand side of the above equations using the inversion formulae \ref{invertedid} yields the antisymmetric part of the identities. 
\smallskip
 
\noindent \emph{\underline{Identities (vi) and (vii)}}$\quad$
These identities are derived by contracting identity \eqref{yzeqn5} with $K_{m}^{IJ} K_{np}^{KL}$. Note that the $F$ terms in this expression arise through the use of identities (iii), which have been used to simplify the expression.
\smallskip

\noindent \emph{\underline{Identity (viii)}}$\quad$
This is obtained by contracting identity \eqref{yzeqn5} with $K^{IJ}_{mn} K_{pq}^{KL}.$

\subsection{Derivation of the identities in table~\ref{so7id4}}
\noindent \emph{\underline{Identity (i)}}$\quad$
These identities are proved by contracting equation \eqref{yzeqn3} with $K_{m}^{IJ} K_{n}^{KL}$.
\medskip

\noindent \emph{\underline{Identity (ii)--(v)}}$\quad$
These are obtained by contracting  equation \eqref{yzeqn3} with $K_{mn}^{IJ} K_{p}^{KL}$.

\subsection{Derivation of the identities in table~\ref{so7id6}}
\noindent \emph{\underline{Identities (i) and (ii)}}$\quad$
The required result is obtained by contracting identities \eqref{yzeqn61} (with the $-$ sign choice) with $K^{m\, IJ}$ and $K^{mn \, IJ}.$
\smallskip

\noindent \emph{\underline{Identities (iii) and (iv)}}$\quad$
Contract identities \eqref{yzeqn62} (with the $-$ sign choice) with $K^{mn \, IJ} K^{p \, KL}.$

\subsection{Derivation of the identities in table~\ref{so7id3}}

It would, at first sight, appear that the identities in table~\ref{so7id3} are most easily derived analogously to the identities in table~\ref{so7id6}, sketched above, using identities \eqref{yzeqn62} except with the + sign choice. However, in fact they can most simply be derived by contracting identities (iii) and (iv) in table~\ref{so7id2} with $\xi_{m}, \zeta_{m}, \xi_{mn}, \zeta_{mn}, F_{m}$ and $F_{mn}$ and using identities (i), (iii) and (iv) from table~\ref{so7id1} and identities (i), (iii) and (iv) from table~\ref{so7id2} to simplify the resulting expressions. Note that the identities must be derived in the order given in table~\ref{so7id3} as earlier identities are used to obtain later ones.

\subsection{Derivation of the identities in table~\ref{so7id5}}
\noindent \emph{\underline{Identity (i)}}$\quad$
We add $4 \xi^{qr} S_{rmn}$ to both sides of equation (iv) in table~\ref{so7id4},
\begin{equation}
  4 \zeta^{r}{}_{q}T_{mnr} + 4 \xi^{r}{}_{q} S_{mnr} - \frac{1}{9} \mathring \eta_{qmnstuv} \zeta^s T^{tuv} = 8
  S^{s}{}_{[qm} \xi_{n]s} - \frac{4}{3}\xi S_{qmn}. 
\end{equation}
Rearranging the above equation, we conclude that $ \zeta^{r}{}_{q}T_{mnr} + \xi^{r}{}_{q} S_{mnr}$ is fully antisymmetric in $\{q,m,n\}$. Hence identity (i).
\smallskip

\noindent \emph{\underline{Identities (ii) and (iii)}}$\quad$
Fully antisymmetrise the indices in identities (iv) and (v) in table~\ref{so7id4}. This leads to a set of simultaneous equations, which can be solved to obtain the result.

\subsection{Derivation of the identities in table~\ref{so7id7}}
\noindent \emph{\underline{Identity (i)}}$\quad$
These are derived by contracting identities (ii) and (iii) of table~\ref{so7id4} with $\xi^{p}$ and $ \zeta^{p}$, respectively and using identities (iii) of table~\ref{so7id2}.  
\smallskip
    
\noindent \emph{\underline{Identity (ii)}}$\quad$
Contract (ii) and (iii) of table~\ref{so7id4} with $F^{p}$ and use identities (i) of table~\ref{so7id3}.
\medskip

\noindent \emph{\underline{Identity (iii)}}$\quad$
These are the most non-trivial identities to prove. We consider the first of the identities, and the other follows from analogous arguments, or simply interchange symmetry. However, before embarking on the proof, we note that contracting (v) in table~\ref{so7id4} with $F_{q}$ and using identity (iii) of table~\ref{so7id5} leads to an equation for the sum of the two equations in (iii) and not on each separately. Therefore, we need another method.

Contract identity (iii) in table~\ref{so7id6} with $\xi_p$. Hence, using identity (i) of table~\ref{so7id4},
\begin{equation} \label{t7iiide1}
\eta_{mnpqrst} F^p \xi^q S^{rst} = 18 S^{q[mn} F^{p]}{}_{q} \xi_{p}.
\end{equation}
In order to find an expression for $S^{q[mn} F^{p]}{}_{q}$ that is amenable to contraction with $\xi_{p}$, we consider
\begin{equation} \label{t7iiiexp}
 \xi^{[m}{}_{q} \zeta^{n}{}_{r} S^{p] qr}.
\end{equation}
This expression can be simplified in two ways. First, we can use identity (viii) of table~\ref{so7id2} to rewrite $\xi^{[m}{}_{[q} \zeta^{n]}{}_{r]}$ and the identities in tables \ref{so7id1} and \ref{so7id2} can be used to simplify expression \eqref{t7iiiexp}. Another way of simplifying the expression is to observe that, from (ii) in table~\ref{so7id4},
\begin{align*}
  \xi^{[m}{}_{q} \zeta^{n}{}_{r} S^{p] qr} =  \xi^{[m|}{}_{q} \zeta^{q}{}_{r} S^{|np]r}.
\end{align*}
 Hence, we can also rewrite expression \eqref{t7iiiexp} using identity (iv) of table~\ref{so7id2}.
We can now equate the two different expressions to derive
\begin{gather}
 27 S^{q[mn} F^{p]}{}_{q} = 2 \xi \zeta S^{mnp} + 2(9 - \zeta^2) T^{mnp} + 3 \zeta^{[m} S^{np]q} \xi_{q} -6 \zeta S^{q[mn} \xi^{p]}{}_{q} - 6 \xi S^{qmn} \zeta^{p}{}_{q} + 12 \zeta T^{q[mn} \zeta^{p]}{}_{q}.
\end{gather}
The required identity can be deduced by substituting the above equation into expression \eqref{t7iiide1} and simplifying using the identities listed in the tables. 

\section{Comparison of stationary points}
 \label{app:comp}

\renewcommand{\theequation}{C.\arabic{equation}}
\setcounter{equation}{0}

In this appendix, we present table~\ref{table:comG2SU4-}, which gives a list of the various tensors used to construct other stationary point uplifts and the associated identities they satisfy.

\begin{table}[H] 
\renewcommand{\arraystretch}{1.4}
\begin{center}
\caption{\label{table:comG2SU4-}}
\scalebox{0.95}{
\begin{tabular}{@{\extracolsep{0 pt}} c c c}
\toprule
Symmetry & SO(8) tensor identities & Associated SO(7) tensor identities \\
\midrule 
G$_2$ & $C^{\pm}_{IJMN} C^{\pm}_{MNKL} = 12 \delta^{IJ}_{KL} \pm 4 C^{\pm}_{IJKL}$ & 
$ \begin{matrix} \xi_{a} \xi_{a} = (21+ \xi)(3- \xi)\\ 
6 \xi_{a b} = (3+\xi) \delta_{ab} - \frac{1}{(3-\xi)} \xi_{a} \xi_{b} \\
6 S_{abe} S_{cde} = 12 \delta^{ab}_{cd} + \eta_{abcdefg} S_{efg} \\
4 S_{[abc} S_{d]ef}= \eta_{abcdgh[e} S_{f]gh}\\ 6 S_{e[ab} S_{cd]f} = \eta_{abcdgh(e} S_{f)gh} \end{matrix} $ \\[4pt]
\midrule
SU(4)$^-$ & $\begin{matrix} Y^{-}_{IJMN} Y^{-}_{MNKL} = 8 \delta^{IJ}_{KL} - 8 F^{- [I}_{[K} F^{- J]}_{L]}\\
 F^{- K}_{I} F^{- J}_{K} =-\delta_I^J\\ Y^{-}_{M IJK} F^{-M}_{L} = Y^{-}_{M [IJK} F^{-M}_{L]} = (Y^{-}_{M [IJK} F^{-M}_{L]})_{-} \end{matrix} $ 
& $ \begin{matrix}   
     K_{a} K_{a} = 1, \qquad K_{a b } K_{b} =0\\ K_{ac} K_{cb} = K_{a} K_{b} - \delta_{ab} \\
     T_{abc} K_{c} = 0\\T_{acd} T_{bcd} = 4(\delta_{ab} - K_{a} K_{b})
    \end{matrix}$
\\[6 pt]
\bottomrule
\end{tabular}
}
\\[8 pt]
\parbox[c]{6.0 in} {\small List of identities satisfied by G$_2$ and SU(4)$^-$ invariant tensors. We use notation where $(X_{IJKL})_{-}$ refers to the anti-selfdual part of tensor $X.$ The SO(7) tensors $\xi, S$ and $T$ are defined according to the general definitions \eqref{xir} and \eqref{Ss}, and $4 K_{a} = F_{IJ} K_{a}^{IJ}, 4 K_{ab} = F_{IJ} K_{ab}^{IJ}$.}
\end{center}
\end{table}

In G$_2$, the single set of tensors $C^{\pm}$ do not close on themselves at the quadratic level, but one can form new tensors from the contraction of $C^{\pm} C^{\mp}$. However, the new SO(7) tensors that can be defined for these objects are related to $\xi$ and $S$ at the quadratic level, hence there is no simplification in doing this.

\section{Choice of SO(3)$\times$SO(3) invariants}
 \label{app:singleset}

\renewcommand{\theequation}{D.\arabic{equation}}
\setcounter{equation}{0}

\newcommand{\fx}{\hbox{\large$\mathfrak{x}$}}
\newcommand{\fS}{{\mathfrak S}}
\newcommand{\fz}{\hbox{\large$\mathfrak{z}$}}
\newcommand{\fT}{{\mathfrak T}}

The metric \eqref{metricalpha} and the 3-form potential \eqref{Aalpha} have been derived using two sets of $\rm SO(3)\times SO(3)$-invariant geometric objects on $S^7$,  namely, $(\xi,\,\xi_m ,\,\xi_{mn},\, S_{mnp})$ and $(\zeta,\,\zeta_m,\,\zeta_{mn},\,T_{mnp})$, that are associated with two sets of (anti-)selfdual $\rm SO(8)$ tensors $Y^\pm_{IJKL}$ and $Z^\pm_{IJKL}$, respectively.  This choice of invariants is crucial for being able to carry out the simplification of  the metric and the 3-form potential in sections~\ref{sec:metric} and \ref{sec:flux} starting with the uplift formulae \eqref{ans:metric} and \eqref{ans:flux}, and also for the  explicit check of the equations of motion in section~\ref{sec:verif-einst-equat}.
 
However, as we have already discussed in section~\ref{sec:harmonics}, one might as well choose to work with a single set of the geometric objects associated with the particular noncompact generator of $E_{7(7)}$ that parametrises a given stationary point. In our case that means setting
\begin{equation}\label{}
\Phi_{IJKL}\eql \cos\alpha\,Y^+_{IJKL}-\sin\alpha\,Z^+_{IJKL}\,,\quad \Psi_{IJKL}\eql \cos\alpha\,Y^-_{IJKL} + \sin\alpha\,Z^-_{IJKL}\,,
\end{equation}
and expressing the solution in terms of the corresponding set of $\rm SO(7)$ tensors
\begin{equation}\label{defofx}
\begin{split}
\fx_{mn}  \eql \cos\alpha \,\xi_{mn}-\sin\alpha \,\zeta_{mn}\,,\quad 
\fx_m   &\eql \cos\alpha \,\xi_m-\sin\alpha \,\zeta_m\,,\quad 
\fx  \eql \cos\alpha \,\xi-\sin\alpha \,\zeta\,,  \\
\fS_{mnp}& \eql \cos\alpha \,S_{mnp} +\sin\alpha\,T_{mnp}\,.
\end{split}
\end{equation}
To do this one may introduce the complementary set of rotated tensors, $\fz_{mn}$, $\fz_m$, $\fz$ and $\fT_{mnp}$, such that
\begin{equation}\label{rotofxi}
\xi_{mn}\eql \cos\alpha\,\fx_{mn}+\sin\alpha \,\fz_{mn}\,,\qquad \zeta_{mn}\eql -\sin\alpha \,\fx_{mn}+\cos\alpha\,\fz_{mn}\,, \qquad \text{etc.}\,.
\end{equation}
After rewriting the solution  in terms of the rotated tensors, one can check   using identities in section~\ref{sec:so7iden} that all terms involving the additional tensors either cancel out or can be rewritten in terms of \eqref{defofx}. 

The calculation is long and, as one might expect, results in more complicated and less symmetric formulae for the metric and the 3-form potential.  The reason for this is that the geometric objects that are being eliminated,   $\fz_{mn}\,,\ldots,\, \fT_{mnp}$,  are replaced by more complex expressions in terms of sums of products of  tensors that are kept.
To illustrate this point, let us consider the warp factor, $\Delta$, given in \eqref{thewarpD}. At the stationary point \eqref{statval},\footnote{Throughout this section we assume that  $c$ and $s$ are set to their stationary point values. Otherwise, there are additional terms proportional to $\fz$ that must be dealt with separately. We have not analysed that case in detail.}
\begin{equation}\label{detcals}
\begin{split}
  \mathcal{X}_2^2 + 2 c^2 \mathcal{X}_2
    \mathcal{Z}_2 + \mathcal{Z}_2^2 + \mathcal{Y} & 
    \eql  20\,\Big[ \Big(\cos(2\alpha)+{\frac{1}{5}} \Big)\,\xi^2-2\sin(2\alpha) \,\xi\zeta-\Big( \cos(2\alpha)-{\frac{1}{5}}\Big)\,\zeta^2\\
    & \hspace{20 pt} -72\sqrt 10\,(\cos\alpha\,\xi-\sin\alpha\,\zeta)+540\\
    &\eql 24\,\fx^2-16\,\fz^2-72\,\sqrt{10}\,\fx+540\\
    & \eql  40\,\fx^2-72\sqrt 10\,\fx+16\,\fx^m\,\fx^n\,\fS_{mpq}\fS_n{}^{pq}+396\,,
\end{split}
\end{equation}
where,  to eliminate $\fz^2$, in the last step we used the fact that
\begin{equation}\label{idxxSS}
\begin{split}
\fx^m\,\fx^n\,\fS_{mpq}\fS_n{}^{pq}& \eql 9-\xi^2-\zeta^2  \eql 9-\fx^2-\fz^2\,,
\end{split}
\end{equation}
which follows from the identities in tables \ref{so7id1}, \ref{so7id2} and \ref{so7id3}.

One may also note that the $\alpha$-dependence in  the first line in \eqref{detcals} is completely removed by rewriting the right hand side in terms of the rotated tensors using \eqref{rotofxi}. Furthermore, the rotated tensors, $\fx_{mn}\,,\ldots,\, \fS_{mnp}$ and $\fz_{mn}\,,\ldots,\, \fT_{mnp}$, satisfy  the same identities  as $\xi_{mn}\,,\ldots\,,S_{mnp}$ and $\zeta_{mn}\,,\ldots\,,T_{mnp}$, respectively, in tables \ref{so7id1}-\ref{so7id7}.  This means that the calculation is precisely the same for all $\alpha$ and thus we may as well set $\alpha=0$. The problem then is simply to rewrite the metric \eqref{metricalpha} and the 3-form potential \eqref{Aalpha} for $\alpha=0$, solely, in terms of $\xi_{mn}\,,\xi_m$ and $S_{mnp}$. With this in mind, we now turn to the metric tensor \eqref{metricalpha}.

It can be shown that one can write all SO(7) tensors appearing in the metric in terms of a small number of fields constructed from $\xi$, $\xi_m$, $\xi_{mn}$ and $S_{mnp}$ only:
\begin{itemize}
\item [(i)]  scalars
\begin{equation}\label{}
\xi\,,\qquad  \Xi   \equiv \xi^m\xi^nS_{mpq}S_n{}^{pq}\,,
\end{equation}
\item [(ii)]  vectors
\begin{equation}\label{}
\xi_m\,,\qquad \Xi_m \equiv   \xi^nS_{mpq}S_n{}^{pq}\,,
\end{equation}
\item[(iii)] symmetric tensors
\begin{equation}\label{}
\mathring g_{mn}\,,\qquad \xi_m\xi_n\,,\qquad \xi_{mn}\,,\qquad \Xi_m\Xi_n\,,
\end{equation}
and
\begin{equation}\label{}
\begin{split}
 \Xi_{mn} & \eql S_{mpq}S_n{}^{pq}\,, \qquad \hspace{22mm}  \widetilde \Xi_{mn} \eql \xi_m{}^p\xi_n{}^qS_{prs}S_{q}{}^{rs}\,,\\
 \Omega_{mn} & \eql \xi^p\xi^qS_{mpr}S_{nq}{}^r\,, \qquad \hspace{15mm} \widetilde \Omega_{mn} \eql \xi^{pq}S_{mpr}S_{nq}{}^r\,,\\
\Lambda_{mn} & \eql \xi^pS^{qr}{}_{(m}\eta_{n)pqrstu}S^{stu}\,, \qquad \quad \widetilde \Lambda_{mn} \eql \xi^{(p}\xi^{w)}{}_{(m} \eta_{n)pqrstu}S_{w}{}^{qr} S^{stu}\,.
\end{split}
\end{equation}
\end{itemize}

Using the identities, one finds that there  are two relations between the symmetric tensors. One is simple
\begin{equation}\label{}
\xi\,\Xi_{mn}\eql 6\,\widetilde\Omega_{mn}\,,
\end{equation}
while the other involves most of the tensors and is quite complicated. We choose the basis of the symmetric tensors by eliminating $\Xi_m\Xi_n$ and $\widetilde\Omega_{mn}$ from the list. 

Now, the metric \eqref{metricalpha} (with $\alpha=0$ and for general $s$ and $c$) is
\begin{equation}\label{}
g_{mn}\eql {\frac{\Delta^2}{36}}\big[ g_0\,\mathring g_{mn}+g_1\,\xi_{mn}+g_2\xi_m\xi_n+g_3\Xi_{mn}+g_4\widetilde \Xi_{mn}+g_5\Omega_{mn}+g_6\Lambda_{mn}+g_7\widetilde\Lambda_{mn}\big]\,,
\end{equation}
where
\begin{equation}\label{}
\begin{split}
g_0&  \eql-6 \sqrt{2} c \xi  s^3-12 \sqrt{2} c \xi  s+\frac{1}{6} s^4 \left(4 \xi ^2+\Xi +126\right)+2
   \left(\xi ^2+27\right) s^2+36\,,\\
g_1 & \eql 2 s \left(18 \sqrt{2} c+\xi  s \left(7 s^2-6\right)\right)\,, \qquad \hspace{28mm} g_2 \eql -\frac{4 s^4}{3}\,,\\
g_3 & \eql -\frac{1}{6} s^2 \left(-36 \sqrt{2} c \xi  s+\left(11 \xi ^2+63\right) s^2+108\right)\,, \qquad
g_4  \eql -42 s^4\,,\\
g_5 & \eql -\frac{2 s^4}{3}\,,\qquad \hspace{9mm} g_6  \eql \frac{5 \xi  s^4}{9}-\sqrt{2} c s^3\,,\qquad \hspace{10mm} g_7  \eql -\frac{2 s^4}{3}\,.\\
\end{split}
\end{equation}

This completes the proof that the metric tensor can be expressed entirely in terms of a single set of geometric objects, $\xi_{mn}\,,\xi_m$ and $S_{mnp}$, together with composite tensors that are built from them. A similar result should also hold  for the 3-form potential. Since the solution written in this form is clearly quite complicated, we will not  discuss this further. It should be  clear at this point that the more symmetric basis of invariant tensors used throughout the paper is a much better choice for doing calculations and that it leads to simpler and more symmetric looking formulae.

\section{Ambient coordinate embedding}
 \label{app:expliciteg}

\renewcommand{\theequation}{E.\arabic{equation}}
\renewcommand{\thetable}{E.\arabic{table}}
\setcounter{equation}{0}

In this appendix, we provide an explicit embedding of the $\mathbb{R}^4$
in $\mathbb{R}^{8}$. We use the following representation of
seven-dimensional $\Gamma$-matrices in terms of Pauli matrices:
\begin{gather}
 \Gamma^{1} =  1 \otimes \sigma^2 \otimes \sigma^1, \qquad  \Gamma^{2}
=  1 \otimes \sigma^2 \otimes \sigma^3, \\
 \Gamma^{3} =  \sigma^2 \otimes \sigma^3 \otimes 1, \qquad  \Gamma^{4}
=  \sigma^1 \otimes 1 \otimes \sigma^2, \\
 \Gamma^{5} =  \sigma^3 \otimes 1 \otimes \sigma^2, \qquad  \Gamma^{6}
=  \sigma^2 \otimes \sigma^1 \otimes 1, \\
 \Gamma^{7} = - \sigma^2 \otimes \sigma^2 \otimes \sigma^2.
\end{gather}
In terms of seven-dimensional $\Gamma$-matrices the SO(8) generators
$\Gamma^{AB}$ are~\footnote{In the expression below we use
$\hat{\Gamma}$ for the SO(8) generators in the spinor representation and
denote the seven-dimensional gamma matrices by $\Gamma$ to avoid confusion.
However, we do not make such a distinction elsewhere.}
\begin{equation}
 \hat{\Gamma}^{a b} = \Gamma^{ab}, \qquad  \hat{\Gamma}^{a 8} = - i
\Gamma^{a}.
\end{equation}

In this representation,
\begin{equation}
 \Gamma^{1234} = - \sigma^3 \otimes \sigma^3 \otimes 1, \qquad
\Gamma^{1235} = \sigma^1 \otimes \sigma^3 \otimes 1, \qquad \Gamma^{45}
= -i \sigma^2 \otimes 1 \otimes 1.
 \label{4g2gid}
\end{equation}
Therefore, we can easily verify that for the embedding given by
\begin{equation}
 m_7 \, x^A=  \{ u^1, u^2, v^3, v^4, -v^1, -v^2, u^3, u^4\}
\end{equation}
the SO(3)$\times$SO(3) invariant scalars, equations \eqref{xisc} and
\eqref{zetasc}, are
\begin{equation}
 \xi = - 3 (u\cdot u - v \cdot v) \qquad \zeta = -6 u \cdot v .
\end{equation}
Furthermore,
\begin{equation}
\delta  x^A = \Gamma^{45}_{AB} \, x^B = \{ v^1, v^2, -u^3, -u^4, u^1, u^2, v^3, v^4\}.
\label{g45rot}
\end{equation}
so the $\alpha$ rotation rotates the $u$ coordinates into the $v$ coordinates, and {\it vice versa}.

\newpage

\bibliographystyle{utphys}
\bibliography{so3}

\end{document}